\documentclass[aps,prx,twocolumn,superscriptaddress, nofootinbib,  longbibliography]{revtex4-2}
\usepackage{graphicx} 
\usepackage{dcolumn}  
\usepackage{bm}    
\usepackage{amssymb}  
\usepackage{amsfonts, amsmath, amssymb, mathtools}
\usepackage{amsmath}

\usepackage{lipsum}
\usepackage[normalem]{ulem}
\usepackage{commath}
\usepackage{xcolor}
\usepackage{comment}
\usepackage[toc,page]{appendix}
\usepackage{booktabs}
\usepackage[normalem]{ulem}
\usepackage{braket} 

\usepackage[binary-units=true]{siunitx}
\usepackage{hyperref}
\usepackage{cleveref}
\RequirePackage{textcase}

\begin{document}

\raggedbottom

\newcommand{\yg}[1]{\textcolor{olive}{#1}}
\newcommand{\tk}[1]{\textcolor{blue}{#1}}
\newcommand{\mc}[1]{\textcolor{orange}{#1}}
\newcommand{\tocheck}[1]{\textcolor{red}{#1}}
\newcommand{\prop}{\,\mathord{\propto}\,}

\title{Demonstration of tripartite cat states in two distinct classes of entanglement}

\author{May Chee Loke}
\thanks{Contributed equally to this work}
\affiliation{Centre for Quantum Technologies, National University of Singapore, Singapore}
\author{Jonathan Schwinger}
\thanks{Contributed equally to this work}
\affiliation{Centre for Quantum Technologies, National University of Singapore, Singapore}
\author{Kehui Yu}
\affiliation{Centre for Quantum Technologies, National University of Singapore, Singapore}
\author{Yingshan Zhang}
\affiliation{Centre for Quantum Technologies, National University of Singapore, Singapore}
\author{Amon M. Kasper}
\affiliation{Centre for Quantum Technologies, National University of Singapore, Singapore}
\author{Tanjung Krisnanda}
\email[Corresponding author: ]{tanjung@nus.edu.sg}
\affiliation{Centre for Quantum Technologies, National University of Singapore, Singapore}
\author{Yvonne Y. Gao}
\email[Corresponding author: ]{yvonne.gao@nus.edu.sg}
\affiliation{Centre for Quantum Technologies, National University of Singapore, Singapore}
\affiliation{Department of Physics, National University of Singapore, Singapore}
\date{\today}

\date{\today}

\begin{abstract}
Entanglement is a cornerstone of quantum mechanics and an essential resource for quantum computation, communication, and metrology. While bipartite entanglement is extensively studied, genuine multipartite entangled states remain largely inaccessible in the continuous-variable domain. Here, we experimentally realize continuous-variable tripartite entangled cat states of the two distinct classes with fundamentally inequivalent properties, namely GHZ-cat and W-cat states, encoded across three microwave resonators coupled to an ancillary superconducting transmon. We develop one-to-all conditional controls on the three oscillators by fully utilizing three energy levels of the transmon to generate these states in a single piece of hardware and employ an efficient subspace tomography protocol for state reconstruction. We obtain measurement-corrected state fidelities of $0.83\pm0.02$ for GHZ-cat and $0.70\pm0.02$ for W-cat, both exceeding the thresholds for genuine multipartite entanglement, with the W-state certification further supported by an optimized fully decomposable witness directly computed with the reconstructed density matrix.
We validate the distinct entanglement classes by experimentally demonstrating that pairwise entanglement is present in the W-cat state but absent in the GHZ-cat state. Our technique for the on-demand generation and characterization of inequivalent classes of tripartite continuous-variable entanglement provides a testbed for studying multipartite bosonic entanglement and its robustness to loss, and a building block for future quantum information processing with multipartite bosonic codes.
\end{abstract}

\maketitle

Entanglement is one of the most striking phenomena in physics and lies at the heart of quantum mechanics. Beyond its central role in experimental tests of quantum foundations~\cite{aspect1999_bell}, it also serves as a key resource for a wide range of quantum technologies, including communication~\cite{ekert1991_quantum, bennett1993_teleporting}, computation~\cite{Jozsa2003}, and metrology~\cite{Giovannetti2004}. While early studies mainly focused on canonical bipartite Bell-type entangled states, increasing attention has recently shifted toward the richer setting of many-body entanglement. Multipartite entangled states offer enhanced communication capacity for quantum teleportation~\cite{ Agrawal2006, karlsson1998_quantum}, enable new forms of distributed quantum information processing~\cite{Horodecki2009}, and provide deeper insight into the behavior of complex quantum many-body systems~\cite{preskill2000_quantum}.

The simplest backdrop for these advantages to manifest is the tripartite case, where genuinely multipartite entangled states fall into two inequivalent classes~\cite{dur2000three}: genuine GHZ-type states, which lack pairwise entanglement, and genuine W-type states, which retain residual bipartite entanglement~\cite{guhne2009entanglement,acin2001classification}. These two classes exhibit distinct physical properties and support different applications. GHZ states are widely used in quantum communication~\cite{pan2000_experimental}, secret sharing~\cite{hillery1999_quantum} and metrology~\cite{kielinski2024_GHZ} due to their strong all-to-all correlations. W~states, by contrast, are particularly valuable for their robustness against particle loss owing to their bipartite entanglement structure, making them useful for quantum error correction~\cite{faist2020_continuous}, teleportation and dense coding~\cite{amico2008_entanglement}.

The strong interest in this direction has motivated numerous experimental demonstrations of GHZ and W states across a variety of platforms, including trapped ions~\cite{roos2004control}, superconducting qubits~\cite{dicarlo2010preparation,neeley2010generation}, and photonic qubits~\cite{eibl2004experimental}.
However, these realizations have largely been confined to discrete-variable (DV) systems. Extending such states to the continuous-variable (CV) regime would open new opportunities for exploring quantum many-body physics~\cite{amico2008entanglement} and probing the quantum-to-classical transition~\cite{frowis2018macroscopic}.
In the single-mode setting, CV systems with redundant bosonic encodings~\cite{gottesman2000encoding,mirrahimi2014dynamically} have emerged as a powerful platform for robust quantum information processing, enabling long-lived quantum memories~\cite{reagor2016quantum}, hardware-efficient quantum error correction~\cite{ofek2016extending,sivak2023real,reglade2024quantum}, and efficient quantum metrology~\cite{pan2025realization,zheng2026quantum}. In the two-mode setting, advances have been made through the realization of Bell-cat states~\cite{wang2016schrodinger,hoshi2025entangling}.
However, extension to the multimode setting presents a significant challenge as it requires not only long coherence times in the underlying physical platform but also precise and robust nonlocal control over many-body interactions in a large Hilbert space. Thus, the genuinely multipartite regime of CV entanglement remains far less explored experimentally, despite its potential to offer capabilities beyond those accessible in DV implementations.

\begin{figure*}
    \centering
    \includegraphics[]{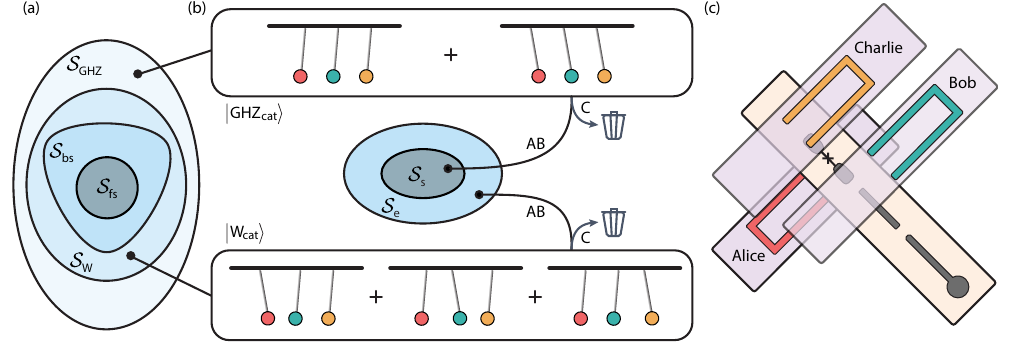}
    \caption{\textbf{Tripartite entanglement landscape and hardware realization.} (a)~Entanglement landscape of a tripartite system. 
    States form a nested hierarchy from fully separable set ($\mathcal{S}_{\text{fs}}$), bi-separable set ($\mathcal{S}_{\text{bs}}$), W set ($\mathcal{S}_{\text{W}}$), and GHZ set ($\mathcal{S}_{\text{GHZ}}$) with $\mathcal{S}_{\mathrm{fs}}\subset \mathcal{S}_{\mathrm{bs}}\subset \mathcal{S}_{\mathrm{W}} \subset \mathcal{S}_{\mathrm{GHZ}}$.
    Genuine multipartite entanglement regime comprises genuine GHZ class (in $\mathcal{S}_{\mathrm{GHZ}}\backslash \mathcal{S}_{\mathrm{W}}$) and genuine W class (in $\mathcal{S}_{\mathrm{W}}\backslash \mathcal{S}_{\mathrm{bs}}$).
    (b)~Entanglement properties of the CV GHZ and W states, $\ket{\text{GHZ}_{\text{cat}}}$ and $\ket{\text{W}_{\text{cat}}}$, illustrated by pendulums oscillating with different phase configurations. 
    Tracing out any one pendulum destroys all remaining entanglement in $\ket{\text{GHZ}_{\text{cat}}}$, resulting in a state belonging to a separable set ($\mathcal{S}_s$). 
    In contrast, the reduced state of $\ket{\text{W}_{\text{cat}}}$ belongs to the entangled set ($\mathcal{S}_e$), highlighting the greater robustness of W-type entanglement.
    (c)~The hardware architecture that enables a one-to-all coupling and conditional controls, where the ancillary transmon is placed on the central chip with readout resonator and Purcell filter, while the three bosonic modes made of tantalum resonators: Alice (red), Bob (turquoise) and Charlie (yellow) are arranged perpendicularly above and below the central chip.
    }
    \label{fig:1}
\end{figure*}


In this work, we experimentally synthesize and characterize two distinct classes of tripartite CV entanglement, in the form of $|\text{GHZ}_\text{cat}\rangle\prop|\alpha,\alpha,\alpha\rangle+|\text{-}\alpha,\text{-}\alpha,\text{-}\alpha\rangle$ and $|\text{W}_\text{cat}\rangle\prop|\alpha,\text{-}\alpha,\text{-}\alpha\rangle+|\text{-}\alpha,\alpha,\text{-}\alpha\rangle+|\text{-}\alpha,\text{-}\alpha,\alpha\rangle$, in three quantum harmonic oscillators hosted in a bosonic circuit quantum electrodynamics (cQED) hardware. We engineer one-to-all controlled unitary operations over the three oscillators and realize on-demand preparation and efficient subspace tomography of GHZ-cat and W-cat states with amplitude $\alpha=1.5$. The states reconstructed via our subspace tomography protocol yield final fidelities of $0.83\pm0.02$ (GHZ-cat) and $0.70\pm0.02$ (W-cat), which are 0.79$\pm$0.02 and 0.64$\pm$0.02 before correction via independently calibrated measurement errors. 
The direct reconstructed state fidelity for GHZ-cat exceeds the threshold for genuine multipartite entanglement~\cite{guhne2009entanglement,acin2001classification}. For the W-cat state, both the measurement-corrected fidelity and fully decomposable witness computed using the Jungnitsch, Moroder, and G\"{u}hne method~\cite{jungnitsch2011taming} on the data without correction confirms GME.
Furthermore, by directly probing selected points in phase space, we also show the violation of Mermin’s inequality~\cite{mermin1990extreme}, which reveals Bell nonlocal correlations and provides an independent verification of genuine multipartite entanglement~\cite{toth2005detecting}. We also experimentally demonstrate that the GHZ-cat state exhibits no bipartite entanglement while the W-cat state retains $64\%$ of its ideal entanglement after a partial trace over one of the modes. 
The techniques shown in this work offer an effective tool for generating distinct classes of tripartite CV entangled states and characterizing their properties. 
Our results mark an important step toward fault-tolerant and error-correctable quantum information processing with multipartite bosonic encodings~\cite{Albert2019PairCatCodes,Denys2023TwoTQutrit}, while also opening a platform for investigating the fundamental structure of entanglement~\cite{zaw2025_witnessing}, its robustness against loss~\cite{brune1996_observing}, and practical witnesses capable of detecting non-Gaussian entanglement~\cite{LinHtoo2024} as well as inequivalent classes of multipartite entanglement~\cite{guhne2009entanglement}.

 \begin{figure*}
    \centering
    \includegraphics[]{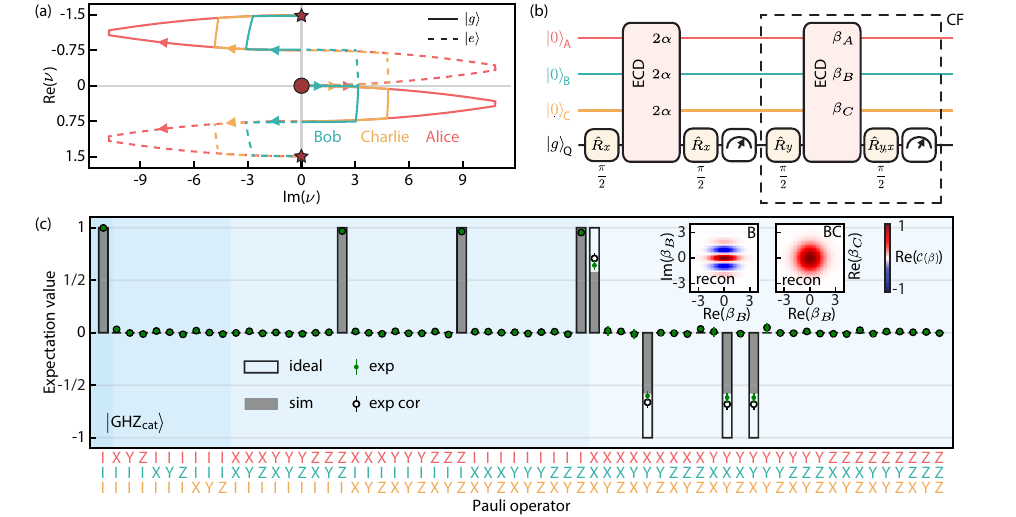}
    \caption{\textbf{GHZ-cat realized using three-mode echoed conditional displacement gate.} 
    (a)~Phase space trajectories of Alice, Bob and Charlie ($\chi_{AQ}/2\pi=57\text{~kHz}, \chi_{BQ}/2\pi=110\text{~kHz}, \chi_{CQ}/2\pi=83\text{~kHz}$) during the implementation of a three mode $ECD(\vec{\beta}=3,3,3)$, starting with vacuum (circle) and ending with coherent states (stars) conditional on the transmon state. 
    Varying the displacement amplitude allows for convenient multimode conditional control without hardware parameter matching. 
    (b)~The preparation of GHZ-cat by initializing the ancillary transmon in equal superposition of $\ket{g}$ and $\ket{e}$ with a transmon rotation followed by a three mode ECD gate with $\alpha=1.5$ to entangle each component of the GHZ-cat to an ancillary state before another transmon rotation is performed to have the GHZ-cat state conditional on the transmon state $\ket{g}$. The GHZ-cat is obtained by post-selecting on ancillary state $\ket{g}$ and is characterized with joint characteristic function (CF) as outlined in the dashed box. 
    (c)~The $64$ Pauli operators measured with joint CF. Our experimental result exhibits the key signature of GHZ state: nonzero three-mode correlations $XXX$, $XYY$, $YXY$, and $YYX$, which are $\approx 62\%$ from the ideal values. 
    Upon tomography-error correction, the value of these correlations see an average improvement factor of $1.10$ (white circles). 
    The inlet represents the real part of the CF of mode B (upon tracing out mode A and C) and of mode B-C upon tracing out mode A, reconstructed from density matrix obtained from post processing the measured Pauli operators. }
    \label{fig:2}
\end{figure*}

In general, the landscape of tripartite quantum states can be organized into four nested sets as illustrated in Fig.~\ref{fig:1}(a): the fully separable set $\mathcal{S}_{\mathrm{fs}}$, the biseparable set $\mathcal{S}_{\mathrm{bs}}$, the W set $\mathcal{S}_{\mathrm{W}}$, and the GHZ set $\mathcal{S}_{\mathrm{GHZ}}$~\cite{guhne2009entanglement, acin2001classification}. The set $\mathcal{S}_{\mathrm{fs}}$ contains mixtures of product states, while $\mathcal{S}_{\mathrm{bs}}\backslash \mathcal{S}_{\mathrm{fs}}$ contains a convex hull of biseparable states with respect to any partition that are not in $\mathcal{S}_{\mathrm{fs}}$. 
Beyond $\mathcal{S}_{\mathrm{bs}}$ lies the regime of genuine multipartite entanglement (GME), in which state preparation necessitates couplings from all parties. Within this regime of interest, $\mathcal{S}_{\mathrm{W}}\backslash \mathcal{S}_{\mathrm{bs}}$ denotes the set of genuine W-type states, while $\mathcal{S}_{\mathrm{GHZ}}\backslash \mathcal{S}_{\mathrm{W}}$ denotes the set of genuine GHZ-type states. At the pure-state level, W-type and GHZ-type states cannot be transformed into one another by stochastic local operations and classical communication (SLOCC)~\cite{dur2000three}.

The most elementary forms of three-mode CV states belonging to the genuine GHZ-type and W-type states simply contain coherent states in each mode:
\begin{eqnarray}
    \ket{\mathrm{GHZ}_{\mathrm{cat}}} &=& \mathcal{N}_{\mathrm{GHZ}} \left(\ket{\alpha,\alpha,\alpha}+\ket{\text{-}\alpha,\text{-}\alpha,\text{-}\alpha}\right), \\
    \ket{\mathrm{W}_{\mathrm{cat}}} &=& \mathcal{N}_{\mathrm{W}} \left(\ket{\alpha,\text{-}\alpha,\text{-}\alpha}+\ket{\text{-}\alpha,\alpha,\text{-}\alpha} +\ket{\text{-}\alpha,\text{-}\alpha,\alpha}\right),\nonumber
    \label{eq:GHZ_W}
\end{eqnarray}
which we refer to as GHZ-cat and W-cat, respectively.
Given the semiclassical nature of coherent states $\ket{\pm\alpha}$, we can consider the three oscillator modes as pendulums and states residing in them as oscillations with opposite phases, as illustrated in Fig.~\ref{fig:1}(b). The distinct phase configurations lead to different classes of entanglement across the three pendulums. In the macroscopic limit of large $\alpha$, where $|\langle \alpha|\text{-}\alpha\rangle| \approx 0$, the normalization factors approach $\mathcal{N}_{\mathrm{GHZ}} \approx 1/\sqrt{2}$ and $\mathcal{N}_{\mathrm{W}} \approx 1/\sqrt{3}$. 

To experimentally synthesize the GHZ-cat and W-cat states effectively, we construct a bosonic cQED architecture that can host three long-lived oscillator modes. Each oscillator takes the form of a superconducting hairpin resonator, as depicted in Fig.~\ref{fig:1}(c). The hairpin resonators are compact and have been shown to exhibit excellent coherence properties~\cite{Ganjam2024}, making them ideal choices for robustly creating and storing tripartite entangled states. Furthermore, we utilize a single transmon to provide the necessary nonlinearity for control and characterization of the three-oscillator states. The transmon is also coupled to a planar readout resonator, together with a Purcell filter, to provide single-shot measurement capability. The interaction between the transmon and the three oscillators is governed by the dispersive coupling,
\begin{eqnarray}
\hat H_{\text{int}}/\hbar&=&\sum_{K=A,B,C}  \frac{\chi_{KQ}}{2}\: \hat k^{\dagger}\hat k \:\hat \sigma_z , \label{EQ_Hbare}
\end{eqnarray}
where $\hat k=\hat a,\hat b,\hat c$ denotes the annihilation operator of the corresponding oscillator $K=A,B,C$, $\chi_{KQ}$ the respective oscillator-transmon dispersive coupling strength, and here the transmon $Q$ is approximated as a two level system with $\hat \sigma_z$ being the Pauli z operator. In general, the always-on dispersive interaction strengths $\chi_{KQ}$ differ for each oscillator mode $K$. This native coupling leads to different rates of phase-space rotations of the corresponding oscillator state conditional on the state of the transmon.
 
Building upon the Hamiltonian in Eq.~(\ref{EQ_Hbare}), together with linear drives of the oscillators and transmon, we implement a one-to-all coupling strategy between the transmon and oscillators using the Echoed Conditional Displacement (ECD), which has been shown to provide universal control on individual oscillators~\cite{eickbusch2022fast}. Here, we achieve efficient joint multimode control using a single transmon by leveraging the two degrees of freedom within an ECD, namely, the linear displacement amplitudes and the wait time over which the oscillator mode undergoes phase-space rotation. We use these two variables to compensate the differences in the coupling strengths between the transmon and each of the oscillators and perform a three-mode ECD of the form of 
\begin{equation}
ECD(\vec \beta)\equiv |e\rangle \langle g|\hat D(\vec \beta/2)+|g\rangle \langle e|\hat D(\text{-}\vec \beta/2),
\end{equation}
where $|g\rangle (|e\rangle)$ denotes the ground (excited) state of the transmon, $\vec \beta=(\beta_A,\beta_B,\beta_C)$ specifies all the complex amplitudes of the coherent states in the three oscillator modes and $\hat D(\vec \beta/2)\equiv \hat D_A(\beta_A/2)\otimes \hat D_B(\beta_B/2)\otimes \hat D_C(\beta_C/2)$ is the three-mode displacement operator.


We illustrate the action of the multimode ECD, $ECD(\vec \beta)$, with $\vec{\beta}=(3,3,3)$, by tracking the phase-space trajectories of all three oscillator modes, as shown in Fig.~\ref{fig:2}(a). By optimizing the individual displacements and wait times of each oscillator, we ensure that in the end all three modes arrive at coherent states of the same amplitude (brown stars) conditional on the state of the transmon. For instance, the oscillator with stronger dispersive coupling to the transmon (e.g. Bob) requires a smaller displacement during the ECD as it rotates faster in phase space at the rate of $\chi_{BQ}$. In contrast, the oscillator with the lowest dispersive coupling needs a larger displacement to compensate for its slower phase accumulation during the wait time. Notably, the three-mode ECD achieves the target multimode dynamics without the stringent hardware requirement of matching all oscillator-transmon coupling strengths, which would otherwise be necessary if the same operation were to be implemented in the strong-dispersive regime. Moreover, the nature of ECD gates allows us to operate effectively with a weak dispersive coupling, which suppresses unwanted higher-order nonlinearities and inter-mode cross-Kerr interactions that would otherwise induce undesired evolutions in the resulting oscillator states. 

The multimode ECD operation, together with single-transmon rotations, directly enables the preparation of GHZ-cat state using the protocol described in Fig.~\ref{fig:2}(b). First, the transmon is prepared in a superposition state $(|g\rangle\,-\,i|e\rangle)/\sqrt{2}$ by applying a transmon $\pi/2$ rotation around the $x$-axis. This is followed by $ECD(2\alpha,2\alpha,2\alpha)$, which entangles the two components of the GHZ-cat state to the ground and excited states of the transmon and yields $(e^{i\phi_{\text{geo}}}|e\rangle |\alpha,\alpha,\alpha\rangle\,-\,i |g\rangle |\text{-}\alpha,\text{-}\alpha,\text{-}\alpha\rangle)/\sqrt{2}$. The parameter $\phi_{\text{geo}}$ arises from a geometric phase induced during the ECD. This is accounted for using independent calibration experiments (Appendix) and omitted in subsequent descriptions for simplicity. Finally, we obtain the desired GHZ-cat state by performing another transmon $\pi/2$ rotation around the $x$-axis followed by a post-selection measurement on the transmon in $|g\rangle$ (Appendix), with ideal (exp) probability of $P_g = 0.5$ ($\approx0.51$).

To characterize our states, we use the three-mode ECD to obtain the joint characteristic function (CF). The CF~\cite{campagne2020quantum} is an informationally complete representation of the state of the system, and is related to the Wigner function through a Fourier transform. With a pre-calibrated $ECD(\beta_A,\beta_B,\beta_C)$, we obtain the joint CF via the sequence presented in the dashed box in Fig.~\ref{fig:2}(b). The real and imaginary parts of the joint CF, $\langle{\hat D(\beta_A,\beta_B,\beta_C)}\rangle$, are retrieved by enacting the last transmon rotation with respect to the y-axis and x-axis, respectively.

Our strategy to perform efficient tomography of the tripartite cat states is based on probing them in the subspace spanned by $\{|\alpha\rangle_A,|\text{-}\alpha\rangle_A\}\otimes \{|\alpha\rangle_B,|\text{-}\alpha\rangle_B\}\otimes \{|\alpha\rangle_C,|\text{-}\alpha\rangle_C\}$, a method previously used to characterize Bell-cat states using selected Wigner-function points~\cite{wang2016schrodinger}.
This is equivalent to a three-qubit tomography on the computational states where $|\alpha \rangle$ ($|\text{-}\alpha \rangle$) represents the computational state $|0\rangle_C$ ($|1\rangle_C$), which is justified as we have approximately orthogonal basis with $|\langle\alpha|\text{-}\alpha\rangle|\approx 0.01$ for $\alpha=1.5$. 
This way, we can retrieve the expectation value of the Pauli operators, where each is retrieved from up to 4 joint CF points. 
For instance, $\langle XXX\rangle=2(\text{Re}(\langle \hat D(2\alpha,2\alpha,2\alpha) \rangle)+\text{Re}(\langle \hat D(\text{-}2\alpha,2\alpha,2\alpha) \rangle)+\text{Re}(\langle \hat D(2\alpha,\text{-}2\alpha,2\alpha) \rangle)+\text{Re}(\langle \hat D(2\alpha,2\alpha,\text{-}2\alpha) \rangle))$. 
For tomographic completeness in the computational space, this subspace tomography requires the retrieval of only $64$ combinations of Pauli operators, just as that of three qubits~\cite{dicarlo2010preparation}. We derive these expectation values from $172$ unique CF points, which is a considerable reduction from a minimum of $\sim10^5$ CF points that would otherwise be required for full-space tomography (Appendix). This framework offers an efficient and robust method to obtain relevant information from highly entangled multimode states within a practical experimental duration and minimizes errors arising from typical parameter drifts. 

Our subspace tomography reveals that the measured Pauli operators (green dots in Fig.~\ref{fig:2}(c)) exhibit features that are consistent with the ideal states (white bars) and close to simulation with decoherence (gray bars).
Importantly, the signature of the GHZ-cat is shown by the nonzero values for the three-mode correlations $XXX$, $XYY$, $YXY$, and $YYX$~\cite{dicarlo2010preparation}.
These quantum correlations, manifested as features far out in the CF, are the most susceptible to oscillators' photon loss, which act as a Gaussian filter centered around the origin~\cite{pan2023protecting}. Therefore, they suffer more significant reductions compared to Pauli operators $III$, $ZZI$, $ZIZ$, and $IZZ$ manifested as features near the origin, which correspond to the photon populations of the oscillators. Equivalently, this main decoherence factor from oscillators' photon loss manifests scrambling of the relative phase of coherent state components, hence inducing a decay of coherence in the computational state elements (e.g.  $|\alpha,\alpha,\alpha\rangle\langle\text{-}\alpha,\text{-}\alpha,\text{-}\alpha|$) during both state preparation and tomography process. In order to accurately analyze the quality of the tripartite entangled state created here, we devise a control experiment to calibrate the errors induced during the tomography step at different locations in phase space. This calibration sequence incurs double the decoherence factor of a single ECD operation. Thus, it provides an independent method to extract the tomography error and allows us to account for it appropriately (Appendix). We present the corrected Pauli data as white circles in Fig.~\ref{fig:2}(c), where the average improvement factor is $1.10$ for the affected three-mode correlations.

\begin{figure}
    \centering
    \includegraphics[]{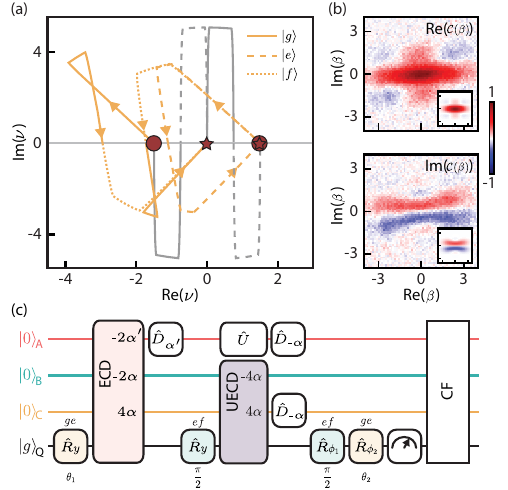}
    \caption{\textbf{Conditional displacement using multiple levels of the transmon and W-cat state preparation protocol.} 
    (a)~An example of the asymmetric phase-space trajectory of mode $C$ during the implementation of $UECD(\beta=(0,0,4\alpha),\alpha_0)$ (yellow) that performs the state transfer of $\ket{g}\ket{\text{-}2\alpha}\mathord{\rightarrow}\ket{f}\ket{0}$, $\ket{e}\ket{2\alpha}\mathord{\rightarrow}\ket{e}\ket{2\alpha}$  and $\ket{f}\ket{2\alpha}\mathord{\rightarrow}\ket{g}\ket{0}$ for $\alpha=3/4$. 
    Circles (stars) indicate the initial (final) coherent states conditional on the transmon state and the gray trajectories represent the ECD used to prepare the initial state for the UECD.
    (b)~The characteristic function of the oscillator $C$ measured after the UECD gate shows good agreement with the simulation of the expected state (inlet). 
    (c)~The protocol to realize W-cat state featuring $ECD(\vec{\beta}=\text{-}2\alpha^{\prime}, \text{-}2\alpha, 4\alpha)$ operated in the g-e subspace and $UECD(\vec\beta=(0,\text{-}4\alpha,4\alpha), \alpha_0)$ in the g-f subspace, together with several unconditional displacement operations and transmon rotations. After post selection of transmon in $\ket{g}$, the final state of the three oscillators is characterized with the same characteristic function (CF) measurement as outlined in Fig.~\ref{fig:2}(b).
    }
    \label{fig:3}
\end{figure}

Furthermore, the tomographically complete set of Pauli measurements allows us to reconstruct the three-qubit density matrix in the computational basis (Appendix) by employing an efficient Bayesian inference procedure~\cite{lukens2020practical}. This allows us to retrieve single and two-mode CFs, as plotted in the inset of Fig.~\ref{fig:2}(c), both in good agreement with the ideal state (Appendix). We also confirm that all two-mode density matrices have entanglement, as quantified by negativity~\cite{zyczkowski1998volume,vidal2002computable}, which are consistent with zero within our uncertainty.
At the three-mode level, we report a fidelity of $0.79\pm 0.02$ and $0.83\pm 0.02$ for the case of experimental data without and with correction, respectively.
Direct calculation of fidelity from the overlap of the measured Pauli operators and the ideal ones agrees with the state reconstruction method (Appendix) to within $5\times10^{-3}$.
This not only concludes that the generated GHZ-cat state possesses GME as the fidelity surpasses the threshold $0.5$, but also is of genuine GHZ class (in the set $\mathcal{S}_{\text{GHZ}}\backslash \mathcal{S}_{\text{W}}$) as it surpasses $0.75$~\cite{guhne2009entanglement,acin2001classification}.
Another validation for GME can be confirmed by violating the Mermin's inequality $M_s\equiv\langle XXX\rangle\,-\,\langle XYY\rangle\,-\,\langle YXY\rangle\,-\,\langle YYX\rangle \le2$. 
Direct evaluation from our measured data reveals $M_s=2.48\pm 0.09$ ($2.7\pm 0.1$) without (with) correction, further affirming the GME~\cite{toth2005detecting} as well as nonlocal~\cite{mermin1990extreme} nature of the GHZ-cat state. 

  
\begin{figure*}
    \centering
    \includegraphics[]{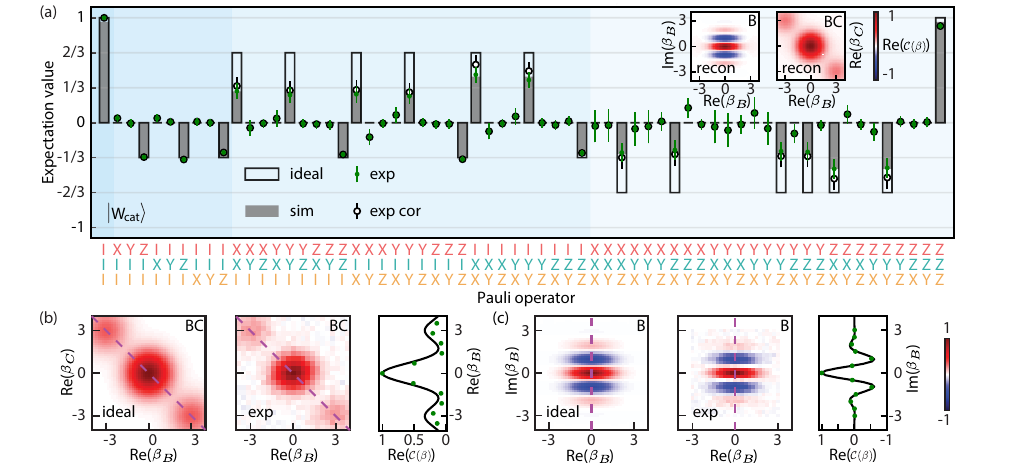}
    \caption{\textbf{W-cat state characterization on three-, two- and single-mode levels.} 
    (a)~The $64$ Pauli operators measured for W-cat state. 
    The entanglement signature of W-cat manifests as nonzero two- and three-mode correlations: $XXI$, $YYI$, $XIX$, $YIY$, $IXX$, $IYY$, $XXZ$, $YYZ$, $XZX$, $YZY$, $ZXX$, and $ZYY$. 
    Upon tomography-error correction, the value of these correlations see an average improvement factor of $1.18$ (white circles).
    The inlet represents the real part of characteristic function of mode B (upon tracing out mode A and C) and of mode B-C upon tracing out mode A, reconstructed from density matrix obtained from post processing the measured Pauli operators. 
    Contrary to the GHZ-cat, the presence of coherence features at $|\beta|=2\alpha$ at two mode level indicates that there is a residual bipartite entanglement. 
    (b)~The ideal and experimental data for the real part of two-mode characteristic function with a plane cut of $Re(\beta_C)$ against $Re(\beta_B)$ and the 1D diagonal cut of the 2D plot. 
    (c)~The ideal and experimental data for the real part of single-mode characteristic function and the 1D vertical cut of the 2D plot.}
    \label{fig:4}
\end{figure*}

We next approach the W-cat state using a similar methodology of entangling the components of the tripartite state to different energy levels of the transmon. In this case, the presence of three components in the superposition necessitates an additional degree of freedom for conditional control of the oscillators, which can be realized by exploiting the readily available multilevel structure of the ancillary transmon. However, while existing ECD protocols are well described for conditional control of an oscillator in the g-e subspace, they do not directly extend to simultaneous conditional control beyond two ancillary levels. For instance, consider the following dispersive Hamiltonian involving multiple levels of the ancillary transmon:
\begin{equation}
        \begin{aligned}
            \hat H_{\text{disp}}/\hbar=\hat a^\dagger \hat a( \Delta_g|g\rangle \langle g|+\Delta_e|e\rangle \langle e|+\Delta_f|f\rangle \langle f| )
        \end{aligned}
        \label{eq:h_disp}
    \end{equation}
where $\Delta_g\equiv \chi_e/2$, $\Delta_e\equiv -\chi_e/2$ and $\Delta_f \equiv \chi_e/2-\chi_f$ denote the transmon state-dependent frequency shifts of oscillator $A$ coupled to transmon $\ket{g}$, $\ket{e}$ and $\ket{f}$, respectively (Appendix). From Eq.~(\ref{eq:h_disp}), it follows that the distribution of the W-cat components onto the three ancillary states cannot be done simultaneously as there is no natural choice of frame that can satisfy $|\Delta_g|=|\Delta_e|=|\Delta_f|$ which is essential for symmetric phase-space evolution. Therefore, we separate the distribution into g-e and g-f subspaces. The g-e conditional control is performed with a multimode ECD, as described in the GHZ-cat preparation. In order to implement a similar construction as the standard ECD in the g-f subspace, the oscillator state should ideally evolve with equal and opposite phases when the transmon is in $|g\rangle$ and $|f\rangle$, and remain stationary when the transmon is in $|e\rangle$. To achieve this, one would position the oscillator drive frequency such that $\Delta_e\rightarrow0$. 
However, this would result in $\Delta_g=\chi_e$ and $\Delta_f=\chi_e-\chi_f$, which give asymmetric trajectories as generally $\Delta_f\ne -\chi_e$ in typical transmon-oscillator devices.
The idea of the Uneven ECD (UECD) gate is to compensate the discrepancies in phase-space evolution induced by the different dispersive shifts by introducing an intentional asymmetry through the oscillator drives.

More specifically, we parametrize the amplitude and phase of each displacement pulses in the oscillators and optimize them to achieve the following effective state-transfer operation for a given initial state $|\alpha_0\rangle$ (Appendix): 
\begin{eqnarray}
    &&UECD(\vec \beta, \alpha_0)\nonumber \\ &&=|f\rangle \langle g|\hat D(\vec \beta/2) + |e\rangle \langle e|\hat I+ |g\rangle \langle f|\hat D(\text{-}\vec \beta/2),
    \label{UECD2}
\end{eqnarray}
where $\vec \beta=(\beta_A,\beta_B,\beta_C)$ denotes the conditional complex displacement vector on the three oscillators and $\hat I$ the identity operator.  

As an example, we illustrate the action of $UECD(\vec\beta=(0,0,4\alpha),\alpha_0)$ on the transmon and oscillator $C$. This operation is designed to simultaneously perform the state transfers $\ket{g}\ket{\text{-}2\alpha}\mathord{\rightarrow}\ket{f}\ket{0}$, $\ket{e}\ket{2\alpha}\mathord{\rightarrow}\ket{e}\ket{2\alpha}$ and $\ket{f}\ket{2\alpha}\mathord{\rightarrow}\ket{g}\ket{0}$. The phase-space trajectory of the oscillator state is shown in Fig.~\ref{fig:3}(a)). The grey lines correspond to the creation of the initial state using $ECD(0,0,4\alpha)$ and the yellow lines describe the evolution of the oscillator state under the $UECD(\vec\beta=(0,0,4\alpha),\alpha_0)$. After the UECD, the oscillator arrives at the final state $\ket{\psi_{C}}=(\sqrt{2}\ket{0}+\ket{2\alpha})/\sqrt{3}$, as marked by the star. We verify this process using CF measurement and our measurement shows excellent agreement with the simulated target state, as shown in Fig.~\ref{fig:3}(b).

Leveraging the multi-level conditional control offered by the UECD, we devise the protocol shown in Fig.~\ref{fig:3}(c) to realize the W-cat state. We start by initializing the transmon in a superposition state $(\sqrt{2}\ket{g}+\ket{e})/\sqrt{3}$ followed by $ECD(\text{-}2\alpha^{\prime},\text{-}2\alpha,4\alpha)$, yielding the following state:
\begin{equation}
    \begin{aligned}
        \ket{\Phi_1}=\sqrt{\frac{1}{3}}
        \ket{g}\ket{\alpha',\alpha,\text{-}2\alpha}
        +\sqrt{\frac{2}{3}}\ket{e}\ket{\text{-}\alpha',\text{-}\alpha,2\alpha},
    \end{aligned}
\end{equation}
where $\alpha^{\prime}=\alpha \exp(i\upsilon)$ is a rotated complex amplitude and $\upsilon$ is chosen such that it compensates the phase space rotation of mode $A$ during the UECD.
After introducing the $\ket{f}$ level of the ancillary transmon with an $R_{y} (\pi/2)$ pulse in the e-f manifold, we perform UECD on mode B and C while mode A undergoes phase space rotation, where the pulse parameters are optimized to obtain the following state:
\begin{equation}
    \ket{\Phi_2}=\frac{1}{\sqrt{3}}
            (\ket{g}\ket{0,\alpha,0}
            +\ket{e}\ket{0,\text{-}\alpha,2\alpha}
            + \ket{f}\ket{2\alpha,\text{-}\alpha,0}).
\end{equation}
After the UECD, unconditional displacements are enacted on mode A and C to arrive at the state where the components of the W-cat state are entangled onto the three ancillary transmon states. We then perform rotations on the e-f and g-e manifolds such that the oscillators' state with the target phase relations is entangled with the transmon $\ket{g}$ state. Finally, a transmon readout allows post-selection for the desired W-cat state, with ideal (exp) probability of $P_g = 1/3$ ($\approx0.34$). 

We perform the same subspace tomography as in the case of GHZ-cat, where the measurement results show expected nonzero values (green dots in Fig.~\ref{fig:4}(a)) for Pauli operators of the ideal W-cat state (white bars), in good agreement with simulation with decoherence (gray bars). The three-mode correlations in the lightest blue shaded area are crucial features for W-cat state, with values expected to be smaller than that of GHZ-cat. Also note that contrary to the GHZ-cat, the W-cat shows nonzero two-mode correlations ($XXI$, $YYI$, $XIX$, $YIY$, $IXX$, $IYY$) that represent coherences corresponding to pairwise entanglement.
A three-mode state reconstruction in the computational space results in $0.64\pm0.02$ ($0.70\pm0.02$) fidelity to the ideal W-cat state for the case without (with) correction. Just as in the GHZ-cat, we attribute the main source of infidelity here to the decay of quantum coherence induced by oscillators' photon loss (Appendix). The relatively lower fidelity compared to the GHZ-cat is due to longer duration of the state preparation protocol, which is $4.9\:\mu$s compared to $3.1\:\mu$s for GHZ-cat and the involvement of the transmon $|f\rangle$ level, which has lower coherence times compared to the $|e\rangle$ level. The measured fidelity is in good agreement with our full error analysis, which takes into account the decoherence effects during each operation (Appendix).

The GME nature of the W-cat state is revealed by exceeding the standard fidelity threshold of $2/3$~\cite{guhne2009entanglement, acin2001classification}, which is surpassed for the measurement-corrected density matrix. To more directly pinpoint whether our state truly possesses GME, we also compute  an optimised fully decomposable witness directly using the reconstructed density matrix without correcting for measurment errors, as introduced by Jungnitsch, Moroder, and G\"{u}hne~\cite{jungnitsch2011taming}. For all the states sampled from the Bayesian distribution, we obtain $\text{tr}(W\rho_i)<0$, with a mean and standard deviation given by $-0.011\pm0.003$, confirming GME of the W-cat state (Appendix). 


To further validate the entanglement structure of the W-cat, we probe the presence of bipartite correlation by directly measuring two-mode CF in selected planes in the phase space. For instance, a plane cut of Bob and Charlie is presented in Fig.~\ref{fig:4}(b). The experimental data closely resemble the ideal CF, with two coherence blobs at the diagonal corners which correspond to the two-mode coherences that are signature of quantum entanglement. The 1D cut plot shows largest deviation at the diagonal coherence blobs, consistent with our error model indicating that the main imperfection arises from the loss of quantum correlations induced by oscillators' photon loss. Explicit computation of entanglement (negativity) from the reconstructed density matrix shows a nonzero value, $0.13\pm0.01$, which is $64\%$ of its ideal value.
Together with the witnessed GME above, this concludes that the W-cat indeed possesses a different entanglement structure (in $\mathcal{S}_{\text{W}}\backslash \mathcal{S}_{\text{bs}}$) compared to the GHZ-cat. Furthermore, the mixed nature of the single mode density matrix is confirmed by measuring the CF of Bob (Fig.~\ref{fig:4}(c)). The measured data agrees very well with ideal CF, shown also in the 1D cut. Intuitively, as the one mode CF does not contain quantum coherence in the computational space, it does not undergo decoherence as severely as the two-mode case (Appendix).

To summarize, we have developed a hardware which enables one-to-all multimode conditional controls to create the two distinct tripartite CV entangled states: GHZ-cat and W-cat states.  The entanglement class of these states is verified via efficient three-mode tomography technique based on mapping the expectation value of Pauli operators to joint three-mode characteristic function points, which are directly measurable within our framework. Our methods prepared the GHZ-cat and W-cat with fidelities of $0.83\pm0.02$ and $0.70\pm0.02$, respectively, and we certified that both states possess GME in their respective classes. GHZ-cat also shows clear violation of the Mermin's inequality with $M_s=2.7\pm 0.1$, further supporting the existence of GME as well as Bell nonlocal correlations. With this, our realization of three-mode GHZ-cat and W-cat states represents a significant extension beyond existing demonstration of two-mode Bell-cat states~\cite{wang2016schrodinger,hoshi2025entangling} towards scalable multimode bosonic entanglement.

The primary factors that limit the fidelities of the GHZ-cat and W-cat states in this study are the relatively short oscillator coherence times of $50-100\:\mu$s. This is particularly detrimental in the multimode setting, as the sensitivity of bosonic-state fidelity to photon loss increases with the number of modes. Using coherence times of 1~ms achievable in state-of-the-art on-chip tantalum resonators reported for the same architecture~\cite{Ganjam2024}, our analysis indicates that the same protocol could prepare the GHZ-cat and W-cat states with fidelities of $0.93\pm0.03$ and $0.88\pm0.03$, respectively. Further improvements on transmon coherence to a moderate $\sim100\:\mu$s would result in fidelities $0.95\pm0.03$ and $0.93\pm0.03$, respectively (Appendix). Nonetheless, the states we created in this work yielded fidelities exceeding the theoretical thresholds required to demonstrate GME in the GHZ and W classes. These results offer concrete experimental insights on the properties of tripartite CV entanglement and serve as a valuable testbed for foundational studies on multipartite quantum systems at the quantum-classical boundary~\cite{frowis2018macroscopic,budroni2022kochen}. Beyond fundamental significance, our demonstration of the joint ECD and UECD operations on three oscillators provides new ingredients for versatile multi-oscillator control. Lastly, our realization of the tripartite cat states with distinct entanglement structures also presents a promising building block for potential multimode quantum error-correction architectures. 

\textbf{Acknowledgment.}
We acknowledge funding support from the Singapore Ministry of Education (A-8004168-00-00) and USydney-NUS Ignition Grants (A-8003773-00-00). J.S, A.K acknowledge the support of the National Quantum Scholarship Scheme (NQSS). We thank Mr.\,Atharv Joshi for the preliminary works on the fabrication of Ta resonators, Dr.\,Steven Touzard, Dr.\,Fumiya Hanamura, Dr.\,Zaw Lin Htoo, Dr.\,Pawe{\l} Cie\'sli\'nski, Prof.\,Valerio Scarani, and Prof.\,Tomasz Paterek for fruitful discussions during the project. We also thank Prof.\,M. Devoret for providing the Josephson Parametric Converter used in the transmon readout process, Dr.\,Xanthe Croot and Dr.\,Yun Li for providing Ta resonator samples that were used to benchmark our in-house fabrication process.

\clearpage
\onecolumngrid

\begin{center}
{\large\bfseries Appendix}

\end{center}

\vspace{1cm}

\setcounter{section}{0}
\setcounter{equation}{0}
\setcounter{figure}{0}
\setcounter{table}{0}

\renewcommand{\thesection}{A\arabic{section}}
\renewcommand{\theequation}{A\arabic{equation}}
\renewcommand{\thefigure}{A\arabic{figure}}
\renewcommand{\thetable}{A\arabic{table}}

\twocolumngrid


\section{Experimental device and system parameters}
\subsection{The device}
\begin{figure}[b]
    \centering
    \includegraphics[]{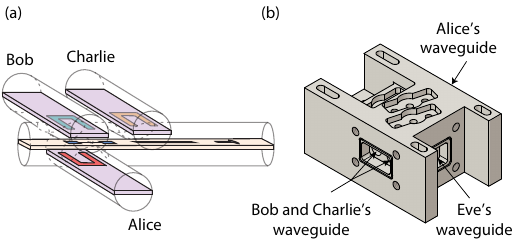}
    \caption{(a) \textbf{Device schematic and package.}
    Schematic of the one-to-all coupling architecture, in which three tantalum hairpin resonators (Alice, Bob, and Charlie) are arranged above and below the central transmon chip, which hosts the ancillary transmon along with its readout resonator and Purcell filter. (b) The high-purity aluminum package that houses the four chips, each in its own waveguide. The chips are horizontally clamped at one end. The clamps are then mounted through screws onto the package.}
    \label{fig:s1}
\end{figure}

The one-to-all coupling architecture is realized with three oscillators coupled to a single ancillary transmon, as shown in Fig. 1(a). The oscillators are on-chip tantalum resonators with a hairpin geometry adapted from Ref.~\cite{Ganjam2024}, using optimized design parameters that balance the different loss mechanisms. The three oscillators (Alice, Bob, and Charlie) are arranged above and below the transmon chip, which hosts the ancillary transmon together with its readout resonator and Purcell filter. The output signal first passes through a quantum limited amplifier in the form of Josephson Parametric Converter (JPC)~\cite{abodo2014_josephson}, followed by a HEMP amplifier at the 4K stage. The design of the high-purity aluminum package that houses the oscillators and the ancillary transmon is shown in Fig.~\ref{fig:s1}(b). The transmon is fabricated from aluminum using standard electron-beam lithography with the bridge-fridge technique and double-angle deposition.

The hairpin tantalum resonators are fabricated by adapting the procedures from Refs.~\cite{Ganjam2024, Crowley2023}. We begin by cleaning the 200\,nm-tantalum film, deposited on sapphire purchased from Star Cryoelectronics, with acetone and methanol. Before photolithography, the wafer is spin-coated with photoresist AZ1512 and baked at $100^{\circ}$C for 1 minute. The film is then patterned using a DMO Microwriter ML3 Pro, followed by development in MF319 for 1 minute and then DI water for 1 minute. After development, the wafer is post-baked at $115^{\circ}$ C for 40\,s. The tantalum film is subsequently dry-etched with inductively coupled reactive ion etching using the Oxford ICP-RIE. We use a chlorine-based dry etch following the recipe of Ref.~\cite{Crowley2023} to ensure sharp, clean edges of the hairpin resonator. The film is etched with the following condition: chlorine and argon flow of 5 sccm each, ICP and HF power of 500\,W and 50\,W respectively, and background pressure of 5.4\,mTorr at $10^{\circ}$ C. After dry etching, the resist is striped off with NMP at $80^{\circ}$ C for 2 hours, followed by cleaning in acetone and methanol before a final thorough chemical processing with piranha for 20 minutes to remove fabrication residue. 

After cleaning, the wafer is diced, and the chips are assembled into the device package. Owing to its extended length, the transmon chip is clamped at both ends to enhance mechanical stability and reduce susceptibility to vibration-induced dephasing. The device is connected to room-temperature control electronics through standard cryogenic microwave wiring, with the input lines appropriately attenuated at multiple temperature stages.

\subsection{Hamiltonian parameters}

\begin{table}[]
\centering
\begin{tabular}{llc}
\toprule
\textbf{Parameter} & \textbf{Description} & \textbf{Value} \\
\midrule
\multicolumn{3}{l}{\textit{Frequencies}} \\
$\omega_A/2\pi$ & Alice                  & 6.485 GHz \\
$\omega_B/2\pi$ & Bob                    & 5.665 GHz \\
$\omega_C/2\pi$ & Charlie                & 5.867 GHz \\
$\omega_Q/2\pi$ & Transmon               & 5.326 GHz \\
$\alpha_Q/2\pi$ & Transmon anharmonicity & 184 MHz   \\
$\omega_R/2\pi$ & Readout                & 7.778 GHz \\
\midrule
\multicolumn{3}{l}{\textit{Dispersive shifts}} \\
$\chi_{AQ}/2\pi$    & Alice, $g$--$e$   & 56(5) kHz  \\
$\chi_{BQ}/2\pi$    & Bob, $g$--$e$     & 109(8) kHz \\
$\chi_{CQ}/2\pi$    & Charlie, $g$--$e$ & 82(4) kHz  \\
$\chi_{f,AQ}/2\pi$ & Alice, $g$--$f$   & 96(2) kHz  \\
$\chi_{f,BQ}/2\pi$ & Bob, $g$--$f$     & 137(5) kHz \\
$\chi_{f,CQ}/2\pi$ & Charlie, $g$--$f$ & 123(1) kHz \\
$\chi_{RQ}/2\pi$    & Readout           & 0.9 MHz    \\
\midrule
\multicolumn{3}{l}{\textit{Higher-order dispersive shifts and self-Kerrs}} \\
$\chi'_{AQ}/2\pi$    & Alice  & $\sim$50 Hz  \\
$\chi'_{BQ}/2\pi$    & Bob    & $\sim$10 Hz \\
$\chi'_{CQ}/2\pi$    & Charlie& --  \\
$K_{AQ}/2\pi$ & Alice  & $\sim$10 Hz  \\
$K_{BQ}/2\pi$ & Bob    & $\sim$10 Hz \\
$K_{CQ}/2\pi$ & Charlie& $\sim$10 Hz \\
\bottomrule
\end{tabular}
\caption{\textbf{Hamiltonian Parameters.}
Hamiltonian parameters of the device. Dispersive shifts are quoted for the transmon $g$--$e$ and $g$--$f$ transitions with each oscillator and with the readout resonator ($\chi_{RQ}$). Numbers in parentheses give the uncertainty in the last digit.}
\label{tab:device_params}
\end{table}

We operate the transmon and oscillators in the weak dispersive regime, which suppresses the Kerr nonlinearity and higher-order terms that would otherwise distort the oscillators' states. Control is enabled by the ECD gate, which amplifies the dispersive interaction to perform fast conditional displacements. The oscillator frequencies are chosen to be well separated, so that each mode remains individually addressable with minimal crosstalk. The measured frequencies and $\chi$s are summarized in Table~\ref{tab:device_params}. We characterize the dispersive Hamiltonian with the out-and-back experiment of Ref.~\cite{eickbusch2022fast}. A large displacement amplifies the weak dispersive terms, a variable wait time lets the oscillator accumulate a transmon-state-dependent phase, and a phase-swept second displacement returns it to vacuum only when its phase is opposite to the accumulated phase of the coherent state. A conditional $\pi$-pulse maps this outcome onto the transmon, allowing us to extract $\chi$.  A variant of the same experiment (see Ref.~\cite{eickbusch2022fast}) yields the cavity's self-Kerr and higher-order dispersive shift $\chi'$, which we find to be a few tens of Hz. We choose the pulses of our conditional displacement gate such that the impact of these terms on the trajectories is negligible.
\subsection{Coherences}

\begin{table}[]
\centering
\begin{tabular}{lccc}
\toprule
 \textbf{Mode} & $T_1$ ($\mu$s) & $T_2^{*}$ ($\mu$s)& $T_2^{\mathrm{echo}}$ ($\mu$s) \\
\midrule
Alice               & 20--230 & --     & --     \\
Bob                 & 40--140   & --     & --     \\
Charlie             & 20--220 & --     & --     \\
Transmon ($g$--$e$) & 40--81   & 16--38 & 30--64 \\
Transmon ($e$--$f$) & 20--42   & 18--26 & --     \\
\bottomrule
\end{tabular}
\caption{\textbf{Coherence times.}
Measured coherence times of the three oscillators and the ancillary transmon. $T_1$ is the energy-relaxation time, $T_2^{*}$ the Ramsey dephasing time, and $T_2^{\mathrm{echo}}$ the Hahn-echo dephasing time; transmon values are given for the $g$--$e$ and $e$--$f$ transitions. Dashes denote quantities not measured, and ranges reflect the spread over the course of the whole experiment which involved many different cool-downs.}
\label{tab:coherences}
\end{table}

As discussed in the main text, the fidelities of the GHZ-cat and W-cat states are primarily limited by the relatively low oscillator coherence times (Table~\ref{tab:coherences}), which fall short of the values ($400\,\mu$s and above) achieved in literature~\cite{Ganjam2024,maiti2025controlling}. We attribute the reduced coherence to several factors: fabrication residues from resist development, dry etching, and dicing; the omission of a buffered-oxide-etch (BOE) treatment of the tantalum oxide layer; and seam loss of the oscillator fields at the Eve-chip seam, which was not fully accounted for during device design. We also observed notable variations in the quality factors of the hairpins in different cool-downs, with $T_1$ ranging from 20-250$\,\mu$s in all the hairpin chips measured in the different iterations of this work. The instances where $T_1<50\,\mu$s are cooldowns where we intentionally couple the hairpins strongly to the drive line to allow strong displacements on them using short drives and test the limits on the gate speed. For the final data presented in the main text, we optimized the balance between the coupling strength and the hairpins' loaded quality factor in order to obtain the final tripartite cat states. With further optimization in both the design and fabrication process, we expect the single photon lifetimes of these hairpin resonators to improve significantly and converge more reliably to the state-of-the-art previously reported in the literature. 

\section{ECD optimization and calibration}
\subsection{Optimization}
Under the idealization of instantaneous displacement pulses, it is straightforward to implement an exact ECD gate. In practice, the finite duration of each pulse causes rotations of the coherent states during the displacement operation distorting their phase-space trajectories. We therefore optimize the four displacement amplitudes, parametrized as $\{\alpha_0 r_0,\ \alpha_0 r_1,\ \alpha_0 r_1,\ \alpha_0 r_2\}$. Given the wanted conditional displacement amplitude $\beta$ and either the wait time or the base displacement amplitude $\alpha_0$, the Nelder-Mead optimiser minimises the cost-function over the semiclassical trajectories (laid out in Ref.~\cite{eickbusch2022fast}) and 
returns the ratios $r_i$ and the unspecified parameter, the wait time or $\alpha_0$. We adapt their cost function with an additional term that enforces convergence to the target $\beta$:
\begin{eqnarray}
        \mathrm{cost} &=& |\alpha_g(T/2)+\alpha_e(T/2)|+|\alpha_g(T)+\alpha_e(T)|\notag\\
        &&+\left|\frac{\alpha_g(T/4)+\alpha_e(T/4)}{2}-\alpha_0\right|\notag\\
        &&+\left|\frac{\alpha_g(3T/4)+\alpha_e(3T/4)}{2}-\alpha_0\right|\notag\\
        &&+2(\beta_\mathrm{current}-|\beta|)^2
        ,\label{eq:cost_function}
\end{eqnarray}
where $T$ is the total duration of the ECD gate, and $\alpha_g$ and $\alpha_e$ are the semiclassical oscillator trajectories for the transmon in the ground and excited state. The base displacement amplitude is $\alpha_0$; $\beta_\mathrm{current}$ is the conditional displacement at the current iteration, given by the magnitude of the difference between $\alpha_g(T)$ and $\alpha_e(T)$, and $\beta$ is the target conditional displacement. We further use the converged trajectories to compute the transmon geometric phase $\theta$ acquired during the ECD, the unconditional oscillator displacement $\gamma$, and the transmon-state-dependent rotation of the oscillator $\phi$, as detailed in Ref.~\cite{eickbusch2022fast}. The cost function drives $\gamma$ to zero and the echo cancels $\phi$, leaving the geometric phase $\theta$, which must be accounted for and corrected in order to obtain the correct final state. We discuss this in more details in Sec.~S4.

\subsection{Vacuum calibration}
\begin{figure}
    \centering
    \includegraphics[]{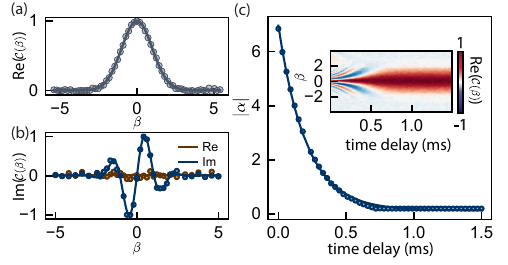}
    \caption{\textbf{ECD calibration and Crosshair measurement.}
    (a) 1D cut of Re(C($\beta$)) of vacuum with DAC amplitude of the ECD displacement adjusted such that $\sigma=1/3$ for ECD($\beta=3$). (b) Crosshair of coherent state $\alpha=1.5$ created by ECD($\beta=3$). (c) Exponential decay of coherent state $\alpha_0$ obtained from a 2D crosshair measurement (inlet) as a function of time delay and fitted to Eq.~(\ref{eq:cavity_t1}), which yields a $T_1$ of 103$\pm{1}\,\mu$s.}
    \label{fig:s2}
\end{figure} 

Once the amplitude ratios $r_i$ and the wait time are fixed by the optimization, the only remaining unknown is the conversion between the played Digital-to-Analog Converter amplitude and the resulting coherent-state size $\alpha_0$. We calibrate this from the characteristic function (CF) of the vacuum, which takes the simple Gaussian form
\begin{equation}
    \mathcal{C}_{|0\rangle}(\beta) = e^{\text{-}|\beta|^2/2}.
    \label{eq:cf_vacuum}
\end{equation}
We measure a one-dimensional cut of $\mathrm{Re}(C(\beta))$ of vacuum while
sweeping the dynamically scaled displacement amplitude. For our base
$\mathrm{ECD}(\beta=3)$ gate, a correctly calibrated DAC amplitude produces
a Gaussian of width $\sigma = 1/3$, as shown in Fig.~\ref{fig:s2}(a). We adjust the DAC amplitude until the condition $\sigma = 1/3$ is satisfied, which simultaneously informs us the exact $\beta$ and the base displacement amplitude $\alpha_0$ in DAC units. Calibrating the CF is thus synonymous with calibrating the ECD displacement amplitude.

\subsection{Geometric phase}
\label{sec:geophase}
The ECD sequence imprints a transmon geometric phase $\theta$ that originates
from the oscillator's phase-space trajectory and takes the form
\begin{equation}
    \theta(t) = -2\int_0^t \mathrm{Re}\!\left[\epsilon^*(\tau)\delta(\tau)\right]d\tau
    + 2\gamma(t)\delta(t),
    \label{eq:geo_phase}
\end{equation}
where $\epsilon$ is the cavity drive, $\delta$ the conditional displacement,
and $\gamma$ the unconditional displacement~\cite{eickbusch2022fast}. Because this phase carries no transmon-state dependence, it cannot be echoed away by the $\pi$ pulse and must be corrected before any subsequent
transmon rotation. For a single ECD, the geometric phase scales with the conditional displacement as $\theta = \theta_0|\beta|^2$, where $\theta_0$ is the phase at $|\beta| = 1$.

\begin{figure}
    \centering
    \includegraphics[]{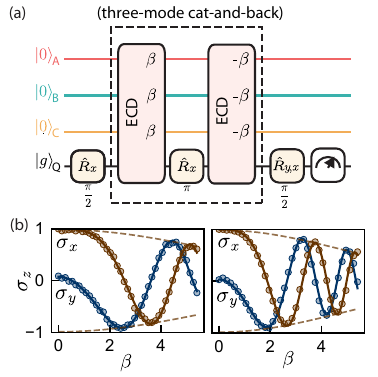}
    \caption{\textbf{Measuring geometric phase with cat-and-back sequence.}
    (a) Three-mode cat-and-back sequence, comprising $\mathrm{ECD}(\beta,\beta,\beta)$, a transmon $\pi$ pulse, and $\mathrm{ECD}(\text{-}\beta,\text{-}\beta,\text{-}\beta)$ (dashed box). The block can be repeated to amplify the phase by inserting an additional transmon $\pi$ pulse between repetitions. (b) Measured $\langle\sigma_x\rangle$ (orange) and $\langle\sigma_y\rangle$ (blue) versus $\beta$ for the single-mode (left) and three-mode (right) cat-and-back. Solid lines are fits to Eq.~\eqref{eq:catback_fit}: the oscillation yields the geometric phase $\theta_0$, while the decaying envelope reflects the coherence loss during the sequence, used to calibrate the error during state characterization
    (Sec.~S6).}
    \label{fig:s3}
    
\end{figure}

We measure $\theta_0$ with a cat-and-back experimental sequence as described in~\cite{eickbusch2022fast} and shown in Fig.~\ref{fig:s3}(a). First, we use a $R_x(\pi/2)$ rotation to prepare the transmon in a superposition state. Subsequently, $\mathrm{ECD}(\beta)$ displaces the oscillator out in phase space, a $R_x(\pi)$ pulse then flips the transmon, and finally, another $\mathrm{ECD}(\text{-}\beta)$ returns the state of the oscillator to vacuum, realizing the effective operation $\sigma_x\,e^{i\theta_0|\beta|^2\sigma_z}$. An additional $R_{y,x}(\pi/2)$ rotation allows us to read out $\langle\sigma_y\rangle$ or
$\langle\sigma_x\rangle$. To extract $\theta_0$, we sweep $\beta$ and fit the measured $\langle\sigma_y\rangle$ and $\langle\sigma_x\rangle$ against
\begin{eqnarray}
    \langle\sigma_x\rangle &=& \cos\!\left(2\theta_0|\beta|^2\right)e^{-\eta\beta^2},\notag\\
    \langle\sigma_y\rangle &=& \sin\!\left(2\theta_0|\beta|^2\right)e^{-\eta\beta^2},
    \label{eq:catback_fit}
\end{eqnarray}
as shown in Fig.~\ref{fig:s3}(b). The additional exponential term $e^{-\eta\beta^2}$ models loss of transmon purity during the sequence as a result of oscillator energy decay as a function of $\beta$. In other words, the features that extend further out in the CF landscape are more susceptible to photon loss~\cite{pan2023protecting}. The resulting decay envelope is used subsequently to calibrate the losses during state characterization (Sec.~S6).

For the multimode tomography, the total geometric phase is the sum of the
single-mode contributions,
\begin{equation}
    \phi_\mathrm{geo}(\vec{\beta}) = \sum_{i=1}^{n}\theta_{0,i}\,|\beta_i|^2,
    \label{eq:geo_multimode}
\end{equation}
where $\vec{\beta} = (\beta_1,\dots,\beta_n)$, $n$ is the number of modes, and $\theta_{0,i}$ is calibrated by a single-mode cat-and-back on cavity $i$. We experimentally verify that the phase from a joint three-mode cat-and-back matches the sum of the individual single-mode phases. Exploiting the real-time computation of the FPGA, we correct $\phi_\mathrm{geo}(\vec{\beta})$ in situ with a virtual-$Z$ rotation on the transmon, applied by adjusting the axis of the final transmon rotation in the CF sequence.

\subsection{Crosshair measurements (Orthogonal 1D cuts of characteristic function)}

Calibrating the ECD gate and preparing our target states both require us to track coherent states in phase space accurately and efficiently. Rather than sampling the full two-dimensional CF, the crosshair technique reconstructs an arbitrary coherent state from only two orthogonal one-dimensional cuts, recovering its full complex amplitude. This makes it our main tool for verifying the ECD gate, calibrating displacements, and measuring the oscillator coherence time.

The CF of a coherent state $\alpha$ is given by
\begin{equation}
    \mathcal{C}_{|\alpha\rangle}(\beta) = \langle\alpha|D(\beta)|\alpha\rangle
    = e^{\text{-}|\beta|^2/2}\,e^{\alpha^*\beta - \beta^*\alpha}.
    \label{eq:cf_coherent}
\end{equation}
Writing the amplitude as $\alpha = a_1 + ia_2$ and tomography displacement as $\beta = b_1 + ib_2$, Eq.~(\ref{eq:cf_coherent}) becomes
\begin{align}
    \mathcal{C}_{|\alpha\rangle}(\beta) = e^{\text{-}|\beta|^2/2}
    [&\cos\!\big(2(a_1 b_2 - a_2 b_1)\big)\nonumber\\
    &+ i\sin\!\big(2(a_1 b_2 - a_2 b_1)\big)].
    \label{eq:cf_crosshair}
\end{align}
The real part of $\alpha$ sets the oscillation frequency along the imaginary displacement axis $b_2$, and the imaginary part along the real axis $b_1$. Therefore, the orthogonal cuts are sufficient to recover all information about $\alpha$. We rely primarily on the imaginary (sine) part, which distinguishes $|+\alpha\rangle$ from $|\text{-}\alpha\rangle$. In Fig.~\ref{fig:s2}(b), we show the crosshair measurement for a coherent state with amplitude $\alpha = 1.5$.

We can also use the crosshair measurement to verify the dispersive coupling $\chi$. We create a coherent state with a single unconditional displacement pulse used in the ECD sequence and probe the resulting state in crosshair experiment. If the expected $\alpha$ is obtained, it confirms that $\chi$ we used in the ECD optimization is correct. Any discrepancies would point to inaccuracies in the values of $\chi$ or the participation of higher order Hamiltonian term like $\chi'$.

Finally, the crosshair provides an alternative measurement of the oscillator's single photon lifetime, $T_1$. The standard approach infers $T_1$ from the decay of the oscillator's photon number population using a number-selective $\pi$ pulse. However, in the weak dispersive regime, the pulse required for sufficient selectivity is typically too long compared to the transmon coherence, introducing significant inaccuracies in the extracted $T_1$. Instead, we displace the oscillator to a large coherent state and track its amplitude with the crosshair as a function of a variable wait time $t$. We then fit the decay to
\begin{equation}
    \alpha(t) = \alpha_0\, e^{-t/2T_1},
    \label{eq:cavity_t1}
\end{equation}
where the factor of two reflects that we measure amplitude rather than energy decay, see Fig.~\ref{fig:s2}(c). This allows us to bypass the issues of the limited selectivity of the transmon pi-pulse and obtain a more truthful estimation of the oscillator lifetime. 

\section{Uneven Echoed Conditional Displacement (UECD)}

The W-cat state is a superposition of three components. In our preparation strategy, each component is mapped onto a separate level of the ancillary transmon and later collected onto $|g\rangle$. Accommodating the third component naturally calls for the transmon's third level $|f\rangle$. The existing ECD framework, however, provides conditional control only within the $g$--$e$ subspace and does not extend to three ancillary levels. In the multilevel dispersive Hamiltonian, Eq.~(4) of the main text, the dispersive shifts $\Delta_g$, $\Delta_e$, and $\Delta_f$ typically cannot simultaneously satisfy $|\Delta_g| = |\Delta_e| = |\Delta_f|$, since the transmon anharmonicity generally makes $2\chi_{e} \neq \chi_{f}$. The phase-space trajectories of the three components are therefore generally different from one another, without any symmetry in the relative phases of the resulting state.

To resolve this issue, we developed the uneven echoed conditional displacement (UECD), which realizes a conditional displacement within the $g$--$f$ subspace while leaving the $|e\rangle$ component undisturbed. The UECD acts as
\begin{equation}
  UECD(\beta, \alpha_0)
  = D\!\left(\tfrac{\beta}{2}\right)|f\rangle\langle g|
  + I\,|e\rangle\langle e|
  + D\!\left(-\tfrac{\beta}{2}\right)|g\rangle\langle f| ,
  \label{eq:uecd-def}
\end{equation}
where the dependence on $\alpha_0$ reflects that the UECD, unlike an ideal conditional displacement, is state-dependent. Its pulses are optimized for one specific configuration of the oscillator state, specified by two reference points. The first, $\alpha_0$, is the point at which the conditional displacement is performed; the $|g\rangle$ and $|f\rangle$ components are displaced from $\alpha_0$ to $\alpha_0 \mp \beta/2$. The second, $\alpha_e(t_0)$, is the initial position of the $|e\rangle$ component. Since the displacement drives act on the oscillator regardless of the transmon state, the $|e\rangle$ component is not automatically left untouched; instead, the pulses are optimized such that the $\ket e$ component returns to itself. Played on any other initial states, the components acquire additional, deterministic rotations, which we derive below and compensate in the W-cat sequence.

\subsection{Pulse parametrization and optimization}
To implement this operation, we build on the ECD drive scheme and compensate the asymmetry in $\Delta_g$ and $\Delta_f$ by freeing up the phase of each displacement pulse. Keeping the ECD's four-displacement structure with a transmon $\pi$ pulse at $t=T/2$, we parametrized our pulses by the set $\{r_1,\phi_1,r_2,\phi_2,r_3,\phi_3\}$, where the two middle pulses share $(r_2,\phi_2)$. An example pulse sequence is shown in Fig.~\ref{fig:s4}(a). 
We choose to operate in the case where all pulses are played in a frame detuned by $\Delta = \chi_{e}/2$ from the dressed oscillator frequencies $\omega_{A,B,C}$, the same frame as the Hamiltonian in Eq.~(2) as well as Eq.~(4) in the main text, the ECD, and the CF tomography. Hence, we can define the transmon state-dependent frequency shifts relative to this frame as $\Delta_g = \chi_{e}/2$, $\Delta_e = -\chi_{e}/2$, and $\Delta_f = \chi_{e}/2 - \chi_{f}$. We determine the six pulse parameters with the Nelder-Mead method by minimizing the cost function
\begin{align}
    \mathrm{cost} =\ & \left(\mathrm{Im}(\alpha_g(T)) - \mathrm{Im}(\alpha_0-\beta/2)\right)^2 \notag\\
    & + \left(\mathrm{Im}(\alpha_f(T)) - \mathrm{Im}(\alpha_0+\beta/2)\right)^2 \notag\\
    & + \left(\mathrm{Re}(\alpha_g(T)) - \mathrm{Re}(\alpha_0-\beta/2)\right)^2 \notag\\
    & + \left(\mathrm{Re}(\alpha_f(T)) - \mathrm{Re}(\alpha_0+\beta/2)\right)^2 \notag\\
    & + |\alpha_e(T) - \alpha_e(t_0)|^2,
    \label{eq:uecd_cost}
\end{align}
whose first four terms drive the $|g\rangle$ and $|f\rangle$ components to their targets $\alpha_0-\beta/2$ and $\alpha_0+\beta/2$, while the last term holds the $|e\rangle$ component fixed at its initial value. The trajectories $\alpha_g$, $\alpha_e$, $\alpha_f$ are computed from the semiclassical evolution of the coherent-state amplitudes,
\begin{equation}
    \alpha_i(t) = e^{-(i\Delta_i+\kappa)t}\left[e^{(i\Delta_i+\kappa)t_0}\alpha_i(t_0)
    - i\int_{t_0}^{t} e^{(i\Delta_i+\kappa)\tau}\varepsilon(\tau)\,d\tau\right],
    \label{eq:uecd_traj}
\end{equation}
where $\Delta_i$ is the drive detuning for the oscillator dressed by the transmon level $i$, $\kappa$ the oscillator relaxation rate, $t_0$ the initial time, and $\varepsilon(t)$ the drive. 

\subsection{Action on a superposition of coherent states}
\label{sec:uecd-superposition}
The UECD sequence is optimised for a particular inital state. However, in our W-state creation protocol, we play the UECD on a superposition of coherent states. Hence, we have to adress how the UECD sequence acts on states different from the optimised for inital state.

 Since Eq.~\eqref{eq:uecd_traj} is linear in the amplitude, the full pulse sequence maps the coherent states on each transmon level $i$ as
\begin{equation}
  \alpha \;\mapsto\; e^{-i\theta_i}\,\alpha + d_i ,
  \label{eq:affine-map}
\end{equation}
Each state experiences a rotation by the angle $\theta_i$ accumulated at the detuning $\Delta_i$, together with a displacement $d_i$ that is independent of the initial state. The angles $\theta_i$ are set by the detunings and the sequence timing alone, while the optimization determines $d_i$. Equation~\eqref{eq:affine-map} therefore determines the action of the UECD on any coherent state.
 
The two echoed branches evolve for one effective half-sequence at $\Delta_g$ and one at $\Delta_f$ and acquire the same total rotation $\Delta\phi = (\Delta_g + \Delta_f)\, T_{\mathrm{tot}}/2$, where $T_{\mathrm{tot}}$ is the time of the entire sequence. By design, we require a state starting at $\alpha_0$ to end at $\alpha_0 \mp \beta/2$, which fixes the drive induced displacment to $d = \alpha_0 \mp \beta/2 - \alpha_0 e^{ -i\Delta\phi}$. A coherent state, different from the optimized on point $\alpha_0$,  starting at an arbitrary $\alpha$ therefore ends at
\begin{equation}
  \alpha e^{-i\Delta\phi} + d
  = \alpha_0 + (\alpha - \alpha_0)\,e^{-i\Delta\phi} \mp \frac{\beta}{2} ,
  \label{eq:offdesign-amplitude}
\end{equation}
i.e., it is first rotated by $-\Delta\phi$ about $\alpha_0$ and then conditionally displaced by $\mp\beta/2$. In operator form, a UECD designed for $\alpha_0$ and played on a coherent state $|\alpha\rangle$ thus realizes
\begin{equation}
  U_{\mathrm{ECD}}(\beta, \alpha_0)\,|\alpha\rangle
  = \mathrm{ECD}(\beta)\,
  D(\alpha_0)\,R(-\Delta\phi)\,D^{\dagger}(\alpha_0)\,|\alpha\rangle ,
  \label{eq:uecd-general}
\end{equation}
where $\mathrm{ECD}(\beta)$ denotes the ideal conditional displacement in the $g$--$f$ subspace and $R(\theta) = e^{i\theta \hat a^\dagger \hat a}$. $D(\alpha_0)\,R(-\Delta\phi)\,D^{\dagger}(\alpha_0)$ implements the rotation about $\alpha_0$. While this rotation vanishes for a standard ECD ($\Delta_g = -\Delta_e$), it has to be accounted for in the UECD whenever $\alpha \neq \alpha_0$.

The $|e\rangle$ component is displaced by the same drives, which act on the oscillator independently of the transmon state, and follows Eq.~\eqref{eq:affine-map} with $\theta_e = \Delta_e T_{\mathrm{tot}}$. The last term of the cost function (Eq.~\eqref{eq:uecd_cost}) requires the point $\alpha_e(t_0)$. The $|e\rangle$ component is thus preserved only at $\alpha_e(t_0)$, which therefore enters the optimization as an input (Table~\ref{tab:uecd_inputs}) and must be set to the actual position of the $|e\rangle$ state in the sequence.

The conditions above constrain only the coherent-state amplitudes. Each transmon level additionally accumulates a geometric phase. These are not corrected within the UECD. They are calibrated and canceled by collecting transmon rotations, as described in Sec.~\ref{sec:Wcat_phase_calibration}.

\begin{figure}
    \centering
    \includegraphics[]{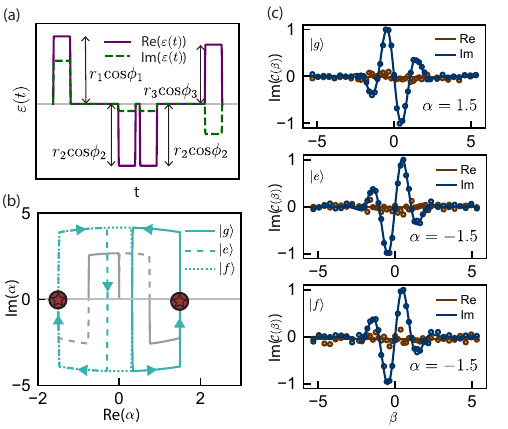}
    \caption{\textbf{UECD state transfer.}
    (a) Pulse sequence and parameters (amplitude ratio $r_i$ and phase $\phi_i$ of UECD state transfer operation. (b) The phase space trajectory of UECD($\vec\beta=(0,-4\alpha,0), \alpha_0$) (turquoise) and ECD for reference (grey). The labels indicate the initial transmon state of the corresponding trajectory. (c) The crosshair measured on oscillator B after running the state transfer operation whereby $\ket{g}\ket{\alpha}\rightarrow\ket{f}\ket{\text{-}\alpha}$, $\ket{e}\ket{\text{-}\alpha}\rightarrow\ket{e}\ket{\text{-}\alpha}$  and $\ket{f}\ket{\text{-}\alpha}\rightarrow\ket{g}\ket{\alpha}$.}
    \label{fig:s4}
\end{figure}

We show the phase space trajectory of the UECD played on mode B during the W-state creation sequence in Fig.~\ref{fig:s4}(b). Starting in the state $\frac{1}{\mathcal N}|g\rangle|\alpha\rangle + |e\rangle|\text{-}\alpha\rangle + |f\rangle|\text{-}\alpha\rangle$, the UECD transfers the different state components as
\begin{align}
    |g\rangle|\alpha\rangle    &\longrightarrow |f\rangle|\text{-}\alpha\rangle, \notag\\
    |e\rangle|\text{-}\alpha\rangle  &\longrightarrow |e\rangle|\text{-}\alpha\rangle, \\
    |f\rangle|\text{-}\alpha\rangle  &\longrightarrow |g\rangle|\alpha\rangle. \notag
\end{align}
We perform crosshair measurements of the oscillator states on $\ket g$, $\ket e$, and $\ket f$ in mode B after both ECD and UECD are played, as shown in Fig.~\ref{fig:s4}(c). We find each component at its expected positions in phase space, which confirms that the sequence has been implemented faithfully on this oscillator. The same crosshair measurements are also carried out in the other two modes, both of which also yield the expected final states. 

\section{GHZ-cat and W-cat creation protocols}
\subsection{GHZ-cat preparation}

The GHZ-cat is prepared with the sequence shown in Fig.~2(b) of the main text. Starting with the transmon in $|g\rangle$ and all three oscillators in vacuum, a $\pi/2$ rotation about the $x$-axis prepares the superposition in the form of
\begin{equation}
    \Psi_1 = \tfrac{1}{\sqrt{2}}\big(|g\rangle|0,0,0\rangle - i\,|e\rangle|0,0,0\rangle\big).
\end{equation}
Then, a three-mode $\mathrm{ECD}(2\alpha,2\alpha,2\alpha)$ conditionally displaces
all three oscillators and entangle the two GHZ-cat components with the transmon
states, i.e.
\begin{equation}
    \Psi_2 = \tfrac{1}{\sqrt{2}}\big(e^{i\phi_\text{geo}}|e\rangle|\alpha,\alpha,\alpha\rangle
    - i\,|g\rangle|\text{-}\alpha,\text{-}\alpha,\text{-}\alpha\rangle\big),
\end{equation}
where $\phi_\text{geo}$ is the geometric phase acquired during the ECD. Finally, a second $\pi/2$ rotation about the $x$-axis recombines the components,
\begin{align}
    \Psi_3 = &0.5|g\rangle\big(|\alpha,\alpha,\alpha\rangle + |\text{-}\alpha,\text{-}\alpha,\text{-}\alpha\rangle\big)\nonumber\\
    &+ 0.5|e\rangle\big(|\alpha,\alpha,\alpha\rangle - |\text{-}\alpha,\text{-}\alpha,\text{-}\alpha\rangle\big),
\end{align}
such that post-selecting on $|g\rangle$ yields the target even-parity GHZ-cat,
$(|\alpha,\alpha,\alpha\rangle + |\text{-}\alpha,\text{-}\alpha,\text{-}\alpha\rangle)/\sqrt{2}$. The
geometric phase $\phi_\text{geo}$ is corrected by a virtual-$Z$ rotation applied together with the second $\pi/2$ pulse, calibrated as described in
Sec.~\ref{sec:geophase}.

\subsection{W-cat preparation}

The W-cat is prepared with the sequence shown in Fig.~3(c) of the main text,
by distributing its three components onto the transmon levels $|g\rangle$,
$|e\rangle$, $|f\rangle$ and then collecting them onto $|g\rangle$. Starting
from the transmon in $|g\rangle$ and the oscillators in vacuum, a transmon rotation
prepares the superposition with ratios,
\begin{equation}
    \Psi_1 = \sqrt{\tfrac{2}{3}}\,|g\rangle|0,0,0\rangle
    + \sqrt{\tfrac{1}{3}}\,|e\rangle|0,0,0\rangle .
\end{equation}
A three-mode ECD entangles the $|g\rangle$ and $|e\rangle$ components with the
oscillators,
\begin{equation}
    \Psi_2 = \sqrt{\tfrac{1}{3}}\,|g\rangle|\alpha',\alpha,\text{-}2\alpha\rangle
    + \sqrt{\tfrac{2}{3}}\,e^{i\phi_{\text{geo},1}}
    |e\rangle|\text{-}\alpha',\text{-}\alpha,2\alpha\rangle ,
\end{equation}
where $\alpha' = \alpha e^{i\upsilon}$ is rotated to compensate for the rotation mode A undergoes while the UECD later acts on the other modes. An unconditional displacement on mode A, then sends its $|e\rangle$ component to vacuum, so that it does not accumulate an unwanted rotation while idling,
\begin{equation}
    \Psi_3 = \sqrt{\tfrac{1}{3}}\,|g\rangle|2\alpha',\alpha,\text{-}2\alpha\rangle
    + \sqrt{\tfrac{2}{3}}\,e^{i\phi_{e,\text{ECD}}}
    |e\rangle|0,\text{-}\alpha,2\alpha\rangle .
\end{equation}
We then introduce the $|f\rangle$ level with a $\pi/2$ pulse on the $e$--$f$
manifold and apply a UECD on modes B and C. Together with further
unconditional displacements, this maps the three W-cat components onto the
three transmon levels,
\begin{equation}
    \Psi_4 = \sqrt{\tfrac{1}{3}}\big(|g\rangle|1\rangle
    + e^{i\phi_1}|e\rangle|2\rangle + e^{i\phi_2}|f\rangle|3\rangle\big),
    \label{eq:Wcat_distributed}
\end{equation}
with $|1\rangle = |\text{-}\alpha,\alpha,\text{-}\alpha\rangle$,
$|2\rangle = |\text{-}\alpha,\text{-}\alpha,\alpha\rangle$,
$|3\rangle = |\alpha,\text{-}\alpha,\text{-}\alpha\rangle$ and $\phi_1$, $\phi_2$ the
geometric phases on the $|e\rangle$ and $|f\rangle$ components. Two further
transmon rotations, on the $e$--$f$ and $g$--$e$ manifolds, collect the three
components onto $|g\rangle$ while equalizing their weights and cancelling
$\phi_1$ and $\phi_2$. Post-selecting on $|g\rangle$ then yields the W-cat,
\begin{equation}
    |W\rangle = \sqrt{\tfrac{1}{3}}\big(|\text{-}\alpha,\alpha,\text{-}\alpha\rangle
    + |\text{-}\alpha,\text{-}\alpha,\alpha\rangle
    + |\alpha,\text{-}\alpha,\text{-}\alpha\rangle\big).
\end{equation}
The geometric phases $\phi_1$ and $\phi_2$ are calibrated as described in
Sec.~\ref{sec:Wcat_phase_calibration}.

The three oscillators play different roles in the sequence, set by the form of their respective target components. Mode A reaches its target state after a single ECD in the $g$--$e$ subspace, while modes B and C require a UECD. A $g$--$f$ exchange on mode B, and an inverse UECD ($\mathrm{UECD}^{-1}$) on mode C that returns the $|g\rangle$ and $|f\rangle$ states to vacuum, see Fig.~3(a) of the main text.  While modes B and C undergo the UECD, mode A idles and rotates. The unconditional displacement after the ECD sends its $|e\rangle$ component to vacuum, leaving only the $|g\rangle$ component, which acquires a phase $\phi_\text{idle} = (\chi_{f} -\chi_{e})\,T_\text{UECD}/2$ over the UECD. We cancel it by playing the preceding ECD on a rotated axis with phase $-\phi_\text{idle}$. In practice, we assign the modes A, B, and C based on the different Hamiltonian parameters and quality factors of each oscillator mode such that the imperfections in the process affect the three modes to a similar extent. 

The optimization inputs for modes B and C to obtain the UECD operation used in the final W-cat data are listed in Table~\ref{tab:uecd_inputs}, and the resulting optimized pulse parameters are presented in Table~\ref{tab:uecd_outputs}. The inputs are a consequence of the coherent state components at the time each UECD is played. For mode B, the $|g\rangle$ and $|f\rangle$ coherent states at $\pm\alpha$ are exchanged while the $|e\rangle$ state idles at $-\alpha$. Choosing the design point $\alpha_0 = 0$, the center of the
pair, keeps the residual rotation of Eq.~\eqref{eq:uecd-general} symmetric between the two coherent states, so that it can be fully absorbed by setting the target conditional displacement of the optimization to $\beta_{opt} = -2\alpha\,(1 + e^{-i\Delta\phi})$. This choice of parameters allows us to implement the UECD with $\beta = -4\alpha$. For mode C, we aim to return the $|g\rangle$ and $|f\rangle$ states to vacuum while the $|e\rangle$ state idles at $2\alpha$. Instead of optimizing for a sequence that directly achieves this, we optimize a UECD with initial state $\ket{0}$ that maps $|g\rangle$ and $|f\rangle$ to the amplitudes of mode C in $\Psi_4$. This implements an UECD with $\beta = -4\alpha$. To achieve the actually wanted operation, an UECD with $\beta = 4\alpha$, we play the inverse of the found sequence.

 In both cases, $\alpha_e(t_0)$ is set to the position of the coherent state $|e\rangle$.

\begin{table}[h]
\centering
\begin{tabular}{lcccc}
\toprule
Mode & $\beta_{opt}$ & $\Delta$ & $\alpha_e(t_0)$ & $\alpha_0$ \\
\midrule
B & $-2\alpha e^{-i\Delta\phi}-2\alpha$ & $\chi_{e}/2$ & $-\alpha$  & $0$ \\
C & $-4\alpha$                  & $\chi_{e}/2$ & $2\alpha$ & $0$ \\
\bottomrule
\end{tabular}
  \caption{\textbf{UECD optimization inputs.} UECD optimization inputs
 for modes B and C. $\beta_{opt}$ is the target conditional displacement, $\Delta$ the drive-frame detuning, $\alpha_e(t_0)$ the fixed point of the $|e\rangle$ trajectory, and $\alpha_0$ the optimization reference point. $\Delta\phi = (\Delta_g + \Delta_f) T_{\mathrm{tot}}/2$ is the residual frame rotation.}
  \label{tab:uecd_inputs}
\end{table}

\begin{table}[h]
\centering
\begin{tabular}{lcc}
\toprule
Parameter & Mode B & Mode C \\
\midrule
Base $\alpha$     & 3.88     & 7.27 \\
$r_1$             & 1.00     & 1.00 \\
$r_2$       & 1.07     & 0.78 \\
$r_3$             & 1.00     & 0.83 \\
$\phi_1$          & $0.000$ & $-0.409$ \\
$\phi_2$ & $0.500$  & $0.018$ \\
$\phi_3$          & $0.000$  & $0.425$ \\
Disp. pulse length (ns)  & 28       & 188 \\
Wait time (ns)    & 824      & 504 \\
\bottomrule
\end{tabular}
\caption{\textbf{Optimized UECD pulse parameters for the W-cat creation.}
Optimized UECD pulse parameters for modes B and C. The two middle
(echo) pulses share their ratio and phase ($r_2=r_3$, $\phi_2=\phi_3$), and phases are in units of $\pi$.}
\label{tab:uecd_outputs}
\end{table}

\subsection{Cavity tomography phase calibration}
\begin{figure}
    \centering
    \includegraphics{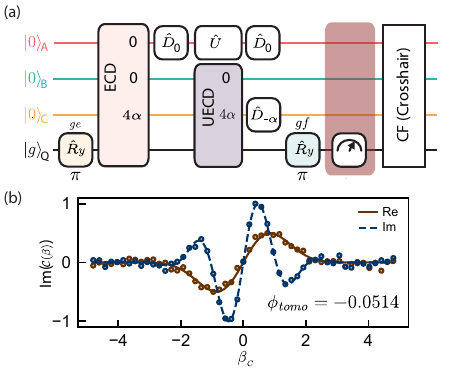}
    \caption{\textbf{Cavity tomography phase calibration.} (a) Pulse sequence, with the post-selection and wait time highlighted (red box). (b) Crosshair
    measurement from which the tomography phase $\phi_\text{tomo}$ is fitted.}
    \label{fig:s6}
\end{figure}

During the post-selection measurement and the subsequent wait time (red box in Fig.~\ref{fig:s6}(a)), each cavity undergoes a phase-space rotation proportional to $\chi t/2$, relative to the measurement frame. We calibrate and correct this rotation by applying a compensating phase to the tomography displacement pulses, which aligns the tomography axis with the rotated state and removes any apparent rotation in the tomography frame.

To calibrate the phase for each mode, we run the state-creation sequence (GHZ- or W-cat), including the post-selection measurement conditioned on $\ket g$. We do this mode by mode, setting the displacement amplitudes on the non under-test mode to zero.  Fitting the crosshair to $\alpha = a_1 + i a_2$ quantifies the rotation the cavity acquired during the post-selection measurement. We align the tomography displacements by applying the same phase. An example of the fitted tomography phase $\phi_\text{tomo}$ is shown in Fig.~\ref{fig:s6}(b).

\section{Tomography in the computational subspace}

\subsection{Mapping the expectation value of Pauli operators to characteristic function points}

Full CF tomography of a three-mode bosonic state requires
extensive sampling of 8-dimensional real and imaginary parts of the CF
\begin{equation}
    \mathcal{C}(\beta_a,\beta_b,\beta_c) = \langle D_a D_b D_c\rangle
    = \mathrm{tr}\!\left(D_a D_b D_c\,\rho\right),
\end{equation}
where $\beta_a$, $\beta_b$, and $\beta_c$ are complex displacement points.
For example, for a coherent state with $\alpha = 1.5$, a truncation of
$N_\text{trunc}=7$ is needed to account for $\geq99\%$ of the total photon population 
per mode. Thus, a three-mode state occupies a Hilbert space of minimal dimension
$N = 7^3$. Sampling the corresponding CF space requires a minimum of $N^2-1\approx 1.2\times10^5$ points, which is both time-consuming and increasingly susceptible to parameter drift. 
Instead, we exploit the near-orthogonality of the coherent states, $|\langle\alpha|\text{-}\alpha\rangle|\approx 0.01$ at $\alpha = 1.5$, which lets us treat $|\alpha\rangle$ and $|\text{-}\alpha\rangle$ as the computational states of a two-dimensional subspace. The three-mode tomography then reduces to measuring the 64 Pauli operators of a ``three-qubit
state"; since the expectation value of each Pauli operator maps to up to four CF points, this amounts to only 172 unique displacement points, considerably less than $1.2\times10^5$.

Under the mapping of $|\alpha\rangle\rightarrow|0\rangle_C,|\text{-}\alpha\rangle\rightarrow|1\rangle_C$, we can construct Pauli operators in the form of
\begin{equation}
\begin{aligned}
  I &= |\alpha\rangle\langle\alpha|+|\text{-}\alpha\rangle\langle\text{-}\alpha|\approx (D_0 + D_0^\dagger)/2 ,\\
  X &= |\alpha\rangle\langle\text{-}\alpha| + |\text{-}\alpha\rangle\langle\alpha|\approx ( D_{2\alpha} + D_{\text{-}2\alpha}),\\
  Y &= i|\text{-}\alpha\rangle\langle\alpha|-i|\alpha\rangle\langle\text{-}\alpha|\approx i( D_{\text{-}2\alpha} - D_{2\alpha}) ,\\
  Z &= |\alpha\rangle\langle\alpha|-|\text{-}\alpha\rangle\langle\text{-}\alpha|\approx s\,i( D_{-i\pi/4\alpha} - D_{i\pi/4\alpha}).
\end{aligned}
\label{eq:pauli}
\end{equation}
where $s = (1/2)\,e^{(\pi/4\alpha)^2/2}$ is a scaling factor that keeps
the expectation value within $\pm 1$. 
The derivation can be confirmed by projecting the operators to the computational subspace. 
For example, $X=(|\alpha\rangle\langle\alpha|+|\text{-}\alpha\rangle\langle\text{-}\alpha|)( D_{2\alpha} + D_{\text{-}2\alpha})(|\alpha\rangle\langle\alpha|+|\text{-}\alpha\rangle\langle\text{-}\alpha|)\approx |\alpha\rangle\langle\text{-}\alpha| + |\text{-}\alpha\rangle\langle\alpha|$.

\begin{figure}
    \centering
    \includegraphics{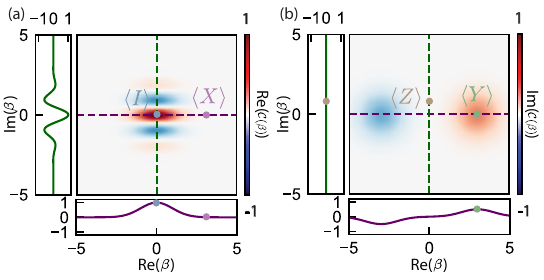}
    \caption{\textbf{Mapping Pauli operators to points in the characteristic function.}
    (a) Real part of the CF of the cat state $|\psi\rangle = \mathcal N(|\alpha\rangle + i|\text{-}\alpha\rangle)$ with $\alpha = 1.5$. $\langle I\rangle$ is obtained by probing the origin, which always returns $1$ regardless of the state, while $\langle X\rangle$ is read at $\mathrm{Re}(\mathcal{C}(\beta = 2\alpha))$. (b)~Imaginary part of the CF of the same state $|\psi\rangle$ (an eigenstate of $Y$), so $\langle X\rangle$ and $\langle Z\rangle$ vanish. $\langle Y\rangle$ is read at the centre of the coherence blob, $\mathrm{Im}(\mathcal{C}(\beta = 2\alpha))$. To visualise $\langle Z\rangle$, consider instead the $Z$ eigenstates $|\pm\alpha\rangle$, whose CFs carry a sinusoidal feature in the imaginary part whose sign at $\beta = i\pi/4\alpha$ distinguishes $|\alpha\rangle$ from $|\text{-}\alpha\rangle$.}
    \label{fig:s7}
\end{figure}

We can thus measure the full Pauli set
$\{I, X, Y, Z\}$ by probing different points of the CF (the expectation values in Eqs.~(\ref{eq:pauli})).
For a single mode, the expectation values are respectively $\text{Re}(\langle D_0\rangle)$, $2\text{Re}(\langle D_{2\alpha}\rangle )$, $2\text{Im}(\langle D_{2\alpha}\rangle )$, and $2s\,\text{Im}(\langle D_{i\pi/4\alpha}\rangle )$.
A visual example for the cat state $|\psi\rangle = \mathcal N\big(|\alpha\rangle + i|\text{-}\alpha\rangle\big)$ is shown in Fig.~\ref{fig:s7}. 
$\langle I\rangle$ is read off at the origin of the real part of the CF; by definition $\mathcal C(0)$ is the trace of the density matrix and therefore equal to $1$. $\langle X\rangle$ and $\langle Y\rangle$ are obtained from the centers of the cat's interference blobs, which appear in either the real or the imaginary part of the CF depending on the phase of the cat. Finally, $\langle Z\rangle$ distinguishes $|\alpha\rangle$ from $|\text{-}\alpha\rangle$ through the sinusoidal feature in the imaginary part of the CF. 


\begin{table}[h]
\centering
\begin{tabular}{cccc}
\toprule
$\Sigma$ & $s$ & $k$ & $\gamma$ \\
\midrule
$I$ & $1/2$ & $0$ & $0$ \\
$X$ & $1$   & $0$ & $2\alpha$ \\
$Y$ & $1$   & $1$ & $-2\alpha$ \\
$Z$ & $\tfrac{1}{2}\,e^{(\pi/4\alpha)^2/2}$ & $1$ & $-i\pi/4\alpha$ \\
\bottomrule
\end{tabular}
\caption{\textbf{Pauli-CF mapping parameters.} The table gives the scaling $s$, the flag $k$ selecting the real or imaginary part, and the location $\gamma$ of the point.}
\label{tab:Pauli_params}
\end{table}

To characterize a three-mode state with this strategy, we express each
three-mode Pauli operator into several CF points. 
First, note that for a single mode, each Pauli operator maps to one CF point, specified by a displacement value and by whether it is taken from the real or imaginary part of the CF. 
We write this into the single expression
\begin{equation}
   \Sigma \approx s\,i^k\big( D_\gamma + (-1)^k D_{-\gamma}\big),
    \label{W:eq:general_1mode}
\end{equation}
where $\Sigma$ is the Pauli operator, $s$ the scaling factor, $k$ selects the
real or imaginary part of the CF, and $\gamma$ the location of the point. The mapping parameters $\{s, k, \gamma\}$ are given in Table~\ref{tab:Pauli_params}.

We generalize
Eq.~\eqref{W:eq:general_1mode} to three modes,
\begin{equation}
\begin{aligned}
\Sigma_1 \Sigma_2 \Sigma_3 
  ={}& s_1 i^{k_1}( D_{\gamma_1} + (-1)^{k_1} D_{-\gamma_1}) \\
  &\otimes s_2 i^{k_2}( D_{\gamma_2} + (-1)^{k_2} D_{-\gamma_2}) \\
  &\otimes s_3 i^{k_3}( D_{\gamma_3} + (-1)^{k_3} D_{-\gamma_3}) \\
  ={}& s_{123} i^{k_{123}}\big( D_{\gamma_1\gamma_2\gamma_3}
      + (-1)^{k_{123}}  D_{-\gamma_1-\gamma_2-\gamma_3} \\
  &+ (-1)^{k_1}  D_{-\gamma_1\gamma_2\gamma_3}
      + (-1)^{k_{23}}  D_{\gamma_1-\gamma_2-\gamma_3} \\
  &+ (-1)^{k_2}  D_{\gamma_1-\gamma_2\gamma_3}
      + (-1)^{k_{13}}  D_{-\gamma_1\gamma_2-\gamma_3} \\
  &+ (-1)^{k_3}  D_{\gamma_1\gamma_2-\gamma_3}
      + (-1)^{k_{12}}  D_{-\gamma_1-\gamma_2\gamma_3}\big),
    \label{W:eq:3ModePauliPoints}
\end{aligned}
\end{equation}
where $s_{123} = s_1 s_2 s_3$, $k_{12} = k_1 + k_2$, $k_{13} = k_1 + k_3$, $k_{23} = k_2 + k_3$, $k_{123} = k_1 + k_2 + k_3$ and
$D_{\gamma_1\gamma_2\gamma_3} = D_{\gamma_1}\otimes D_{\gamma_2}\otimes D_{\gamma_3}$. 
For any three-mode Pauli operator, the last line of
Eq.~\eqref{W:eq:3ModePauliPoints}, together with the single-mode parameters of Table~\ref{tab:Pauli_params}, fixes the displacement points, scaling factors,
and the signs of the sum.
The displacement operators are paired so that, depending on $\{k_1,k_2,k_3\}$, they yield
either the real or imaginary part of the corresponding displacement operator.
Every three-mode Pauli expectation
value is therefore mapped to up to four points in the 8D CF.
For example, $\langle III\rangle=\text{Re}(\langle D_0D_0D_0\rangle)$, while $\langle XXX\rangle=2(\text{Re}(\langle D_{2\alpha}D_{2\alpha}D_{2\alpha}\rangle)+\text{Re}(\langle D_{-2\alpha}D_{2\alpha}D_{2\alpha}\rangle)+\text{Re}(\langle D_{2\alpha}D_{-2\alpha}D_{2\alpha}\rangle)+\text{Re}(\langle D_{2\alpha}D_{2\alpha}D_{-2\alpha}\rangle))$.

The parity vector $\vec k = (k_1,k_2,k_3)$ takes eight values, giving eight
groups of Pauli operators, each expressed as a sum of up to four
CF points, as listed in Table~\ref{tab:k_group_mapping}.
The naive count of $64\times4 = 256$ points reduces to $172$ unique ones
because some operators collapse to fewer points.

\begin{table*}[t]
\centering
\renewcommand{\arraystretch}{1.4}
\begin{tabular}{ccc}
\toprule
$\mathbf{k}$ & Pauli operators & Characteristic-function mapping \\
\midrule
$000$ & $\{I,X\}^{\otimes 3}$ &
$s_{123}\big[2\text{Re}\langle D_{\gamma_1,\gamma_2,\gamma_3}\rangle
+2\text{Re}\langle D_{-\gamma_1,\gamma_2,\gamma_3}\rangle
+2\text{Re}\langle D_{\gamma_1,-\gamma_2,\gamma_3}\rangle
+2\text{Re}\langle D_{\gamma_1,\gamma_2,-\gamma_3}\rangle\big]$ \\
\addlinespace
$110$ & $\{Y,Z\}\otimes\{Y,Z\}\otimes\{I,X\}$ &
$s_{123}\big[-2\text{Re}\langle D_{\gamma_1,\gamma_2,\gamma_3}\rangle
+2\text{Re}\langle D_{-\gamma_1,\gamma_2,\gamma_3}\rangle
+2\text{Re}\langle D_{\gamma_1,-\gamma_2,\gamma_3}\rangle
-2\text{Re}\langle D_{\gamma_1,\gamma_2,-\gamma_3}\rangle\big]$ \\
\addlinespace
$101$ & $\{Y,Z\}\otimes\{I,X\}\otimes\{Y,Z\}$ &
$s_{123}\big[-2\text{Re}\langle D_{\gamma_1,\gamma_2,\gamma_3}\rangle
+2\text{Re}\langle D_{-\gamma_1,\gamma_2,\gamma_3}\rangle
-2\text{Re}\langle D_{\gamma_1,-\gamma_2,\gamma_3}\rangle
+2\text{Re}\langle D_{\gamma_1,\gamma_2,-\gamma_3}\rangle\big]$ \\
\addlinespace
$011$ & $\{I,X\}\otimes\{Y,Z\}\otimes\{Y,Z\}$ &
$s_{123}\big[-2\text{Re}\langle D_{\gamma_1,\gamma_2,\gamma_3}\rangle
+2\text{Re}\langle D_{-\gamma_1,\gamma_2,\gamma_3}\rangle
+2\text{Re}\langle D_{\gamma_1,-\gamma_2,\gamma_3}\rangle
-2\text{Re}\langle D_{\gamma_1,\gamma_2,-\gamma_3}\rangle\big]$ \\
\addlinespace
$111$ & $\{Y,Z\}^{\otimes 3}$ &
$s_{123}\big[2\text{Im}\langle D_{\gamma_1,\gamma_2,\gamma_3}\rangle
+2\text{Im}\langle D_{-\gamma_1,\gamma_2,\gamma_3}\rangle
+2\text{Im}\langle D_{\gamma_1,-\gamma_2,\gamma_3}\rangle
+2\text{Im}\langle D_{\gamma_1,\gamma_2,-\gamma_3}\rangle\big]$ \\
\addlinespace
$100$ & $\{Y,Z\}\otimes\{I,X\}\otimes\{I,X\}$ &
$s_{123}\big[2\text{Im}\langle D_{-\gamma_1,-\gamma_2,-\gamma_3}\rangle
+2\text{Im}\langle D_{-\gamma_1,\gamma_2,\gamma_3}\rangle
+2\text{Im}\langle D_{-\gamma_1,\gamma_2,-\gamma_3}\rangle
+2\text{Im}\langle D_{-\gamma_1,-\gamma_2,\gamma_3}\rangle\big]$ \\
\addlinespace
$010$ & $\{I,X\}\otimes\{Y,Z\}\otimes\{I,X\}$ &
$s_{123}\big[2\text{Im}\langle D_{-\gamma_1,-\gamma_2,-\gamma_3}\rangle
+2\text{Im}\langle D_{\gamma_1,-\gamma_2,-\gamma_3}\rangle
+2\text{Im}\langle D_{\gamma_1,-\gamma_2,\gamma_3}\rangle
+2\text{Im}\langle D_{-\gamma_1,-\gamma_2,\gamma_3}\rangle\big]$ \\
\addlinespace
$001$ & $\{I,X\}\otimes\{I,X\}\otimes\{Y,Z\}$ &
$s_{123}\big[2\text{Im}\langle D_{-\gamma_1,-\gamma_2,-\gamma_3}\rangle
+2\text{Im}\langle D_{\gamma_1,-\gamma_2,-\gamma_3}\rangle
+2\text{Im}\langle D_{-\gamma_1,\gamma_2,-\gamma_3}\rangle
+2\text{Im}\langle D_{\gamma_1,\gamma_2,-\gamma_3}\rangle\big]$ \\
\bottomrule
\end{tabular}
\caption{\textbf{Three-mode Pauli mappings grouped by the parity vector $\mathbf{k}=(k_1,k_2,k_3)$.}
Pauli operators are grouped by the values of $k_i$ from the single-mode
mappings in Table~\ref{tab:Pauli_params}, where $k(I)=k(X)=0$ and
$k(Y)=k(Z)=1$. Operators within a group share the same CF reconstruction formula and differ only through the displacement values
$\gamma_i$ and scaling factors $s_i$.}
\label{tab:k_group_mapping}
\end{table*}

\subsection{Calibrating the W-cat geometric phases}
\label{sec:Wcat_phase_calibration}
Once the three W-cat components are mapped onto the transmon levels, they
carry relative geometric phases $\phi_1$ and $\phi_2$. In the GHZ case an
analogous phase was removed with a cat-and-back, but the $|f\rangle$ level in
the W-cat sequence rules out such a Ramsey-like measurement. Instead, we read
the phases directly from Pauli expectation values, using the
CF mapping established above.

After post-selection, the oscillator state carries the two phases,
\begin{equation}
    |\Phi\rangle = \tfrac{1}{\sqrt{3}}\big(
    |\text{-}\alpha,\alpha,\text{-}\alpha\rangle
    + e^{i\phi_1}|\text{-}\alpha,\text{-}\alpha,\alpha\rangle
    + e^{i\phi_2}|\alpha,\text{-}\alpha,\text{-}\alpha\rangle\big).
    \label{eq:Wphase_state}
\end{equation}

To calibrate $\phi_1$ and $\phi_2$, we first identify which Pauli operators
carry information about them. Expanding the expectation value of a Pauli
operator $\hat P$ on $|\Phi\rangle$, the cross terms between the three
components carry the relative phases,
\begin{equation}
\begin{aligned}
\langle\Phi|\hat P|\Phi\rangle = \tfrac{1}{3}\big[
  &\langle\text{-}\alpha,\alpha,\text{-}\alpha|\hat P|\text{-}\alpha,\alpha,\text{-}\alpha\rangle \\
  &+ e^{-i\phi_1}\langle\text{-}\alpha,\text{-}\alpha,\alpha|\hat P|\text{-}\alpha,\alpha,\text{-}\alpha\rangle \\
  &+ e^{-i\phi_2}\langle\alpha,\text{-}\alpha,\text{-}\alpha|\hat P|\text{-}\alpha,\alpha,\text{-}\alpha\rangle \\
  &+ e^{i\phi_1}\langle\text{-}\alpha,\alpha,\text{-}\alpha|\hat P|\text{-}\alpha,\text{-}\alpha,\alpha\rangle \\
  &+ \langle\text{-}\alpha,\text{-}\alpha,\alpha|\hat P|\text{-}\alpha,\text{-}\alpha,\alpha\rangle \\
  &+ e^{i(\phi_1-\phi_2)}\langle\alpha,\text{-}\alpha,\text{-}\alpha|\hat P|\text{-}\alpha,\text{-}\alpha,\alpha\rangle \\
  &+ e^{i\phi_2}\langle\text{-}\alpha,\alpha,\text{-}\alpha|\hat P|\alpha,\text{-}\alpha,\text{-}\alpha\rangle \\
  &+ e^{i(\phi_2-\phi_1)}\langle\text{-}\alpha,\text{-}\alpha,\alpha|\hat P|\alpha,\text{-}\alpha,\text{-}\alpha\rangle \\
  &+ \langle\alpha,\text{-}\alpha,\text{-}\alpha|\hat P|\alpha,\text{-}\alpha,\text{-}\alpha\rangle \big].
\end{aligned}
\label{eq:Wphase_expansion}
\end{equation}
An operator thus depends on $\phi_1$, $\phi_2$, or $\phi_2-\phi_1$ whenever it
connects the corresponding pair of components of the W-cat. The operators which are sensitive to each geometric phase factor are listed in Table~\ref{tab:phase_pauli}.

\begin{table}[h]
\centering
\begin{tabular}{ccc}
\toprule
$\phi_i$ & $\propto \cos\phi_i$ & $\propto \sin\phi_i$ \\
\midrule
$\phi_1$          & $IXX,\,IYY,\,ZXX,\,ZYY$; & $IXY,\,IYX,\,ZXY,\,ZYX$ \\
$\phi_2$          & $XXI,\,YYI,\,XXZ,\,YYZ$; & $XYI,\,YXI,\,XYZ,\,YXZ$ \\
$\phi_2 - \phi_1$ & $XIX,\,YIY,\,XZX,\,YZY$; & $XIY,\,YIX,\,XZY,\,YZX$ \\
\bottomrule
\end{tabular}
\caption{\textbf{Phase-sensitive Pauli operators.} Operators whose
expectation value on $|\Phi\rangle$ varies as the cosine or sine of each
geometric phase.}
\label{tab:phase_pauli}
\end{table}

\begin{figure}
    \centering
    \includegraphics{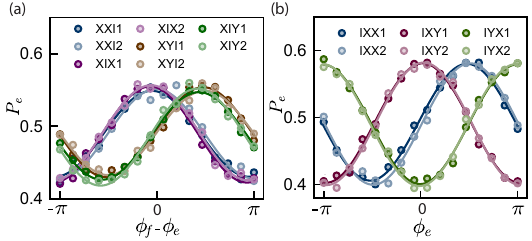}
    \caption{\textbf{W-cat geometric-phase calibration.} Characteristic-function
    points of the phase-sensitive Pauli operators versus the swept virtual-$Z$
    phase: (a) the $\phi_2$-sensitive operators as $\phi_{f}-\phi_e$ of the $e$--$f$
    rotation is varied, and (b) the $\phi_1$-sensitive operators as $\phi_{e}$
    of the $g$--$e$ rotation is varied. Solid lines are sinusoidal fits, from
    which we extract the virtual-$Z$ angles that cancel $\phi_1$ and $\phi_2$.}
    \label{fig:5}
\end{figure}

We calibrate $\phi_1$ and $\phi_2$ by running the W-cat sequence while varying
the virtual-$Z$ phase $\phi_{e}$ ($\phi_{f}-\phi_e$) of the $g$--$e$ ($e$--$f$)
$\pi/2$ rotation and measuring the CF at the displacement
points $\mathcal{C}(\beta_{\hat P})$ of the relevant Pauli operators. We first set
$\phi_{f}-\phi_e=0$, sweep $\phi_{e}$, and monitor the CF
points of the $\phi_1$-sensitive operators $\langle IXX\rangle$ and
$\langle IXY\rangle$, extracting $\phi_{e}$ from a sinusoidal fit. We then fix
$\phi_{e}$ at this value, sweep $\phi_{f}-\phi_e$, and monitor the
$\phi_2$-sensitive operators $\langle XXI\rangle$, $\langle XIX\rangle$,
$\langle XYI\rangle$, and $\langle XIY\rangle$. Finally, we extract $\phi_{f}-\phi_e$ from a
second sinusoidal fit, as shown in Fig.~\ref{fig:5}.

\subsection{State reconstruction}

The measurements of 64 different Pauli operators are tomographically complete; they can be used to reconstruct the density matrix of the three oscillators' states in the computational space.

Suppose we define $\vec \rho$ and $\vec X$ as the vectorized version of the density matrix of the state to be characterized (column-wise stacking) and the list of the corresponding measured Pauli operators (observables), respectively. 
They are linearly related as
\begin{equation}
    \vec X = M \vec \rho,\label{EQ_tomolineqn}
\end{equation}
where $M$ is the measurement matrix.
In our case, $M$ is a $64\times64$ matrix, where each of its rows is composed of the elements of the corresponding $8\times 8$ Pauli operator stacked row-wise.

Given the measurement results (the vector $\vec X$ with 64 elements), we solve the density matrix by inverting Eq.~(\ref{EQ_tomolineqn}) together with Lagrange multiplier, where the constraint is $\text{tr}(\rho)=1$.
We denote the resulting solution as $\vec \rho_{\text{est}}$, giving a density matrix $\rho_{\text{est}}$ with unit trace.
However, the density matrix $\rho_{\text{est}}$ might not be physical; it might contain negative eigenvalues caused by experimental errors in $\vec X$.
To get a physical estimated density matrix, given $\rho_{\text{est}}$, we employ an efficient Bayesian inference method proposed in Ref.~\cite{lukens2020practical}. 
The method gives us a statistically accurate distribution, which is used to compute the distribution of estimated density matrices $\{\rho_i\}$, or quantities of interest directly.
In this work, we retrieve a distribution of fidelities $\{F_i\}$ of the estimated density matrices to the corresponding ideal state $|\psi \rangle$ with
\begin{equation}
    F_i=\langle \psi|\rho_i|\psi \rangle.
\end{equation}
In the main text, we report the average fidelity with the standard deviation from the distribution.
We also note that one can directly compute fidelity from the measured Pauli operators via
\begin{eqnarray}
    F&=&\langle \psi|\rho|\psi \rangle \nonumber \\
    &=&\frac{1}{8}\sum_{j=1}^{64} \langle P_j \rangle_{\text{exp}} \langle P_j \rangle_{\text{ideal}},
\end{eqnarray}
where $\langle P_j \rangle_{\text{exp}}$ is the measured value of the $j$th Pauli operator and $\langle P_j \rangle_{\text{ideal}}$ is that of the ideal value.
The fidelities computed this way differ from those reported in the main text by less than $5\times10^{-3}$.

\begin{figure}
    \centering
    \includegraphics[]{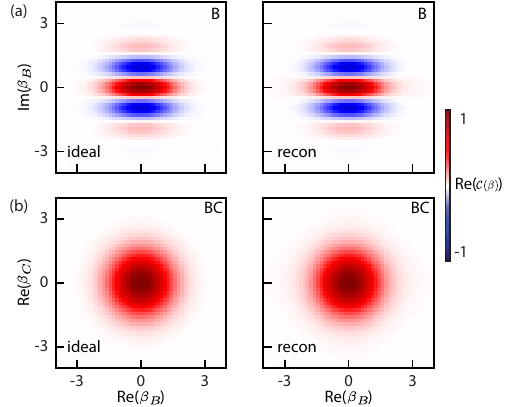}
    \caption{\textbf{Marginals of the GHZ-cat state.} (a) The CF of mode $B$ after tracing out the GHZ-cat state with respect to mode $A$ and $C$ in the ideal case (left) and reconstructed from experimental data (right). (b) The CF of mode $B$-$C$ after tracing out the GHZ-cat state with respect to mode $A$ in the ideal case (left) and reconstructed from experimental data (right).}
    \label{fig:GHZ1m2m_ideal_recon}
\end{figure}

Next, we can also compute the Bayesian mean estimate $\rho_{\text{BME}}$, which is the average of the estimated density matrices from the distribution.
This density matrix $\rho_{\text{BME}}$ has dimension $8\times 8$ (defined in the computational space). 
Upon the transformation $|0\rangle_C (|1\rangle_C)\rightarrow |\alpha\rangle (|\text{-}\alpha\rangle)$, we have the estimated density matrix in the larger Hilbert space for the three oscillator modes.
Now we can perform partial trace on this density matrix with respect to mode $A$ and $C$, and compute the CF of the density matrix of mode $B$.
For the case of GHZ-cat, this result is plotted in Fig.~\ref{fig:GHZ1m2m_ideal_recon}(a) (right panel), which is nearly identical to the ideal case (left panel).
Similarly, by performing partial trace with respect to mode $A$, we end up with the density matrix for mode $B-C$, whose CF is plotted in Fig.~\ref{fig:GHZ1m2m_ideal_recon}(b) (right panel) in the plane cut of $\text{Re}(\beta_B)$ vs $\text{Re}(\beta_C)$.
One can see that the reconstructed CF agrees well with that of the ideal state (left panel).
The case of W-cat state is presented in the inlet of Fig.~4(a) in the main text.

\subsection{Fully decomposable witness for W-cat state}

A standard way to reveal GME for a reconstructed W state $\rho$ is by using the witness $W=(2/3)\mathbb{I}-|\psi_W\rangle\langle\psi_W|$~\cite{guhne2009entanglement, acin2001classification} and having $\text{tr}(W\rho)<0$.
In other words, GME is revealed by having the state fidelity $\langle \psi_W|\rho|\psi_W\rangle >2/3$.
In our case, this is true for reconstructed state with correction where the fidelity is $0.70\pm0.02$, but not without correction where the fidelity is $0.64\pm 0.02$.

However, one can use an optimised fully decomposable witness introduced in Ref.~\cite{jungnitsch2011taming} by Jungnitsch, Moroder, and G\"{u}hne.
Here, a semidefinite programming is used to compute $\min_W \text{tr}(W\rho)$, where $W$ is optimised with constraints such that $\text{tr}(W\rho)\ge 0$ for all states in the set $\mathcal{S}_{\text{bs}}$.
For the W-cat state reconstruction without correction, all the states sampled from the Bayesian distribution show $\text{tr}(W\rho_i)<0$, with a mean and standard deviation given by $-0.011\pm0.003$. 
Therefore, we conclude that the experimental data without correction also reveals GME with $100$\% confidence.

\section{Error analysis}

In this section we will detail the error analysis that we develop to explain the imperfections for both the GHZ-cat and W-cat states. The analysis includes independent treatment of a few error sources: transmon energy decay (characterized by its lifetimes $T_1^{ge}$ and $T_1^{ef}$ in the g-e and e-f subspaces, respectively), transmon dephasing (characterized by its lifetimes $T_{\phi}^{ge}$ and $T_{\phi}^{ef}$ in the g-e and e-f subspaces, respectively), and photon losses in the oscillators (characterized by the lifetimes $T_{A1}$, $T_{B1}$, and $T_{C1}$ for mode $A$, $B$, and $C$, respectively).
We will also independently investigate the errors during state preparation and characterization procedures. This way, it gives us deep insight into how much a particular error channel contributes to a particular process in the protocol.

We will first start with preliminary details of error model for each individual system in section~\ref{App_EA_pre}. Here, we will show state transformations under each error channel that will be used in later sections. Section~\ref{App_EA_pauli} presents how the errors in the three-qubit subspace (computational space) translate to the Paulibars for both GHZ-cat and W-cat states. Errors for state preparation and characterization will then be presented in sections~\ref{App_EA_stateprep} and \ref{App_EA_statechar}, respectively.
Section~\ref{App_EA_allerror} will present the accumulation of all errors and comparison with experimental data. Finally, section~\ref{App_EA_controlexp} explains the control experiments we conducted to calibrate out the errors during state characterization.

\subsection{Preliminary details}\label{App_EA_pre}

\subsubsection{Transmon energy decay}

In our work, we populate the transmon up to three energy levels: the ground state $|g\rangle$, first excited state $|e\rangle$, and second excited state $|f\rangle$. 
We performed experiments to retrieve transmon lifetimes $T_1^{ge}$ and $T_1^{ef}$, which characterize the decay of excitation from $|e\rangle$ to $|g\rangle$ and from $|f\rangle$ to $|e\rangle$, respectively. 

The effect of energy decay on the transmon state is captured by the Lindblad master equation 
\begin{equation}
    \dot \rho = \sum_n L_n \rho L_n^{\dagger} -\frac{1}{2} (L_n^{\dagger}L_n \rho + \rho L_n^{\dagger}L_n),
    \label{EQ_Linblad}
\end{equation}
where $L_n$ denotes the jump operator corresponding to the error channel.
For the case of energy decay described above, we take the jump operators as $L_1 = \sqrt{1/T_1^{ge}}\sigma_{ge}$ and $L_2 = \sqrt{1/T_1^{ef}}\sigma_{ef}$ with $\sigma_{ge}\equiv |g\rangle \langle e|$ and $\sigma_{ef}\equiv |e\rangle \langle f|$ being the lowering operators in the g-e and e-f subspaces, respectively.

The transformation of the transmon state for a time $t$, element by element, after solving Eq.~(\ref{EQ_Linblad}) with jump operators $\{L_n\}=\{L_1,L_2\}$, is as follows:
\begin{eqnarray}
    |g\rangle \langle g|&\rightarrow&|g\rangle \langle g|, \nonumber \\
    |e\rangle \langle e|&\rightarrow&|e\rangle \langle e|(1-\frac{t}{T_1^{ge}})+|g\rangle \langle g|\frac{t}{T_1^{ge}}, \nonumber \\
    |f\rangle \langle f|&\rightarrow&|f\rangle \langle f|(1-\frac{t}{T_1^{ef}})+|e\rangle \langle e|\frac{t}{T_1^{ef}}, \nonumber \\
    |g\rangle \langle e|&\rightarrow&|g\rangle \langle e|e^{-t/2T_{1}^{ge}}, \nonumber \\
    |g\rangle \langle f|&\rightarrow&|g\rangle \langle f|e^{-t/2T_{1}^{ef}}, \nonumber \\
    |e\rangle \langle f|&\rightarrow&|e\rangle \langle f|e^{-t(1/2T_{1}^{ge}+1/2T_{1}^{ef})},\label{EQ_QT1}
\end{eqnarray}
where the solutions for the diagonal elements are valid to first order in $t/T_1^{ef}, t/T_1^{ge}$.
The population in f-level decays with a factor $1-t/T_1^{ef}$, and the decay part (with a factor $t/T_1^{ef}$) goes to e-level. 
This process similarly happens for the decay from e-level to g-level.
The energy decays also induce dephasing, i.e., decay for the off-diagonal elements.
The g-e decay affects $|g\rangle \langle e|$ element with a factor $e^{-t/2T_{1}^{ge}}$, e-f decay affects $|g\rangle \langle f|$ element with a factor $e^{-t/2T_{1}^{ef}}$, and the element $|e\rangle \langle f|$ is affected by both decays resulting in a factor $e^{-t(1/2T_{1}^{ge}+1/2T_{1}^{ef})}$.

\subsubsection{Transmon pure dephasing}

Pure dephasing affecting the g-e and e-f subspaces are modeled by jump operators: $L_3 = \sqrt{\gamma_e}|e\rangle \langle e|$ and $L_4 = \sqrt{\gamma_f}|f\rangle \langle f|$. 
These are sufficient to model the decay of all the off-diagonal elements of the density matrix.
The decay rate $\Gamma_{ij}$ for off-diagonal elements can be written as follows
\begin{eqnarray}
    \Gamma_{ge}\equiv \frac{1}{T_{\phi}^{ge}}&=&\frac{\gamma_e}{2} \nonumber \\
    \Gamma_{gf}&=&\frac{\gamma_f}{2} \nonumber \\
    \Gamma_{ef}\equiv \frac{1}{T_{\phi}^{ef}}&=&\frac{\gamma_e+\gamma_f}{2}, 
\end{eqnarray}
where the last term is because $\Gamma_{ef}$ is affected by both jump operators.
This means that for the jump operators, we will use $\gamma_e=2/T_{\phi}^{ge}$ and $\gamma_f=2/T_{\phi}^{ef}-2/T_{\phi}^{ge}$.
By solving Eq.~(\ref{EQ_Linblad}) for a time $t$ with jump operators given by $\{L_n\}=\{L_3,L_4\}$, we have the following transformation
\begin{eqnarray}
    |g\rangle \langle e|&\rightarrow&|g\rangle \langle e|e^{-t/T_{\phi}^{ge}}, \nonumber \\
    |g\rangle \langle f|&\rightarrow&|g\rangle \langle f|e^{-t(1/T_{\phi}^{ef}-1/T_{\phi}^{ge})}, \nonumber \\
    |e\rangle \langle f|&\rightarrow&|e\rangle \langle f|e^{-t/T_{\phi}^{ef}},\label{EQ_QTphi}
\end{eqnarray}
where the diagonal elements are constant and the transformation governing the $|g\rangle \langle f|$ element is induced. 

Note that from Eqs.~(\ref{EQ_QT1}) and Eqs.~(\ref{EQ_QTphi}), the relation between pure dephasing and total dephasing within our model is given by
\begin{eqnarray}
    \frac{1}{T_2^{ge}}&=&\frac{1}{T_{\phi}^{ge}}+\frac{1}{2T_{1}^{ge}}\nonumber \\
    \frac{1}{T_2^{ef}}&=&\frac{1}{T_{\phi}^{ef}}+\frac{1}{2T_{1}^{ef}}+\frac{1}{2T_{1}^{ge}},
\end{eqnarray}
which we use to extract $T_{\phi}^{ge},T_{\phi}^{ef}$ from experimentally measured $T_{2}^{ge},T_{2}^{ef},T_{1}^{ge},T_{1}^{ef}$.

\subsubsection{Oscillator's photon loss}

For the case of the oscillators, instead of presenting the state transformation on the Fock basis, we will present the effect of decoherence on coherent states and their superpositions, which is more relevant to our work.
The jump operators corresponding to photon loss can be written as $L_5=\sqrt{1/T_{A1}}\:\hat a$, $L_6=\sqrt{1/T_{B1}}\:\hat b$, and $L_7=\sqrt{1/T_{C1}}\:\hat c$ for oscillators $A$, $B$, and $C$, respectively. 

For a single oscillator, say Alice, a coherent state $|\alpha \rangle$ will decay to a smaller coherent state $|\alpha e^{-t/(2T_{A1})}\rangle$ in time $t$, under the dynamics of Eq.~(\ref{EQ_Linblad}) with $L_5$ as the jump operator. However, this decoherence is not that significant in our case as the oscillator's lifetime $T_{A1}$ is much longer than the time scales in our protocols. For example, for $t=4.9\:\mu$s, $e^{-t/(2T_{A1})}\approx 0.98$.

A more significant error manifests if the oscillator is in a superposition of coherent states, say $|\psi\rangle_A=\mathcal{N}(|\alpha\rangle +|\beta \rangle)$, where $\mathcal{N}$ is the normalization factor.
In this case, the state will decay into a mixed state, with the following transformations
\begin{eqnarray}
    |\alpha\rangle \langle\alpha|&\rightarrow & |\alpha^{\prime}\rangle \langle\alpha^{\prime}|\nonumber \\
    |\beta\rangle \langle\beta|&\rightarrow & |\beta^{\prime}\rangle \langle\beta^{\prime}| \nonumber \\
    |\alpha\rangle \langle\beta|&\rightarrow & |\alpha^{\prime}\rangle \langle\beta^{\prime}|e^{-2(\delta_{\alpha \beta}/2)^2(1-e^{-t/T_{A1}})},\label{EQ_cat_decay}
\end{eqnarray}
where $\alpha^{\prime}=\alpha e^{-t/(2T_{A1})}$, $\beta^{\prime}=\beta e^{-t/(2T_{A1})}$, and $\delta_{\alpha \beta} = |\alpha-\beta|$.
Note that the ``off-diagonal" element $|\alpha\rangle \langle\beta|$ here not only decays in amplitude but also with a factor that is exponentially worse for the case of large superpositions where the separation $\delta_{\alpha \beta}$ is high.
In the main text, we refer to this as decay of coherence (as it causes decay of the ``off-diagonal" element) induced by photon loss.
This is the root of main decoherence in our work as superpositions of coherent states are created during the protocols.

In our work, we will ignore the decay of state amplitudes as the effect is negligible, and focus mainly on the coherence decay factor. This also means that a single mode mixed state of the form $p_1 |\alpha\rangle\langle\alpha |+p_2|\text{-}\alpha\rangle \langle \text{-}\alpha |$ (with $p_1+p_2=1$) will experience negligible decay as it does not contain coherence element, as confirmed by the exemplary case of mode $B$ in Fig.~4(c) in the main text.

The infinitesimal version of Eqs.~(\ref{EQ_cat_decay}) for a state $|\psi(t)\rangle_A=\mathcal{N}(|\alpha_t\rangle+|\beta_t\rangle)$ at time $t$ evolving for $\Delta t$ then reads
\begin{eqnarray}
    |\alpha_t\rangle \langle\alpha_t|&\rightarrow & |\alpha_t\rangle \langle\alpha_t|\nonumber \\
    |\beta_t\rangle \langle\beta_t|&\rightarrow & |\beta_t\rangle \langle\beta_t| \nonumber \\
    |\alpha_t\rangle \langle\beta_t|&\rightarrow & |\alpha_t\rangle \langle\beta_t|e^{-2(\delta_{\alpha_t \beta_t}/2)^2\Delta t/T_{A1}},\label{EQ_cat_decay_inf}
\end{eqnarray}
where $\delta_{\alpha_t \beta_t}=|\alpha_t-\beta_t|$ and the decoherence only affects the off-diagonal element.
As the amplitudes change in time when we apply our gates (the values of which can be computed using the semiclassical approach~\cite{eickbusch2022fast}), the decay factor of the quantum coherence will accumulate as
\begin{equation}
    f_A= e^{\int-2(\delta_{\alpha_t \beta_t}/2)^2/T_{A1} \: dt}.\label{EQ_cat_decay_inf_sum}
\end{equation}

A useful extension of Eq.~(\ref{EQ_cat_decay_inf_sum}) for three mode where all jump operators $\{L_5,L_6,L_7\}$ are present is especially useful in this work.
For a three mode state of the form $|\psi\rangle =\mathcal{N}(|\alpha_{At},\alpha_{Bt},\alpha_{Ct}\rangle + |\beta_{At},\beta_{Bt},\beta_{Ct}\rangle)$, the decaying factor for the off-diagonal element $|\alpha_{At},\alpha_{Bt},\alpha_{Ct}\rangle \langle\beta_{At},\beta_{Bt},\beta_{Ct}|$ is $\tilde f \equiv f_Af_Bf_C$ with 
\begin{eqnarray}
    f_A&=& e^{\int-2(\delta_{\alpha_{At} \beta_{At}}/2)^2/T_{A1} \: dt}\nonumber \\
    f_B&=& e^{\int-2(\delta_{\alpha_{Bt} \beta_{Bt}}/2)^2/T_{B1} \: dt}\nonumber \\
    f_C&=& e^{\int-2(\delta_{\alpha_{Ct} \beta_{Ct}}/2)^2/T_{C1} \: dt}.\label{EQ_3Mcoherence_decay_factor}
\end{eqnarray}
It is important to note that the decaying factor depends on the \emph{separation} between the ket $|\alpha_{At},\alpha_{Bt},\alpha_{Ct}\rangle$ and bra $\langle\beta_{At},\beta_{Bt},\beta_{Ct}|$ of the off-diagonal element. For states with more coherent state components (e.g. W-cat), all off-diagonal elements will decay concurrently, and the decaying factor for each element can be computed in similar manner.

\subsection{Errors mapped to Pauli bars}\label{App_EA_pauli}

Both the GHZ-cat and W-cat live in a space spanned by $\{|\alpha\rangle_A,|\text{-}\alpha\rangle_A\}\otimes \{|\alpha\rangle_B,|\text{-}\alpha\rangle_B\}\otimes \{|\alpha\rangle_C,|\text{-}\alpha\rangle_C\}$. 
When decoherence is taken into account, the value for each element in this space will decay from the ideal one. 
The error analysis that we develop retrieves these values after decoherence. 
In this section we translate these values directly to the expectation value of Pauli operators (Paulibars), such that the latter can be directly compared with experimental data.
For ease of notation and comparison to the three-qubit case, we will use computational notation where $|\alpha\rangle \rightarrow |0\rangle$ and $|\text{-}\alpha\rangle \rightarrow |1\rangle$.

We start with the GHZ-cat, where the state after the application of a decoherence channel takes the form
\begin{eqnarray}
    \rho &=& \eta_1|000\rangle \langle 000|+\eta_2|000\rangle \langle 111|\nonumber \\
    &&+\eta_3|111\rangle \langle 000|+\eta_4|111\rangle \langle 111|,\label{EQ_GHZdecayform}
\end{eqnarray}
where $\eta$s denote the coefficients after decoherence.
The expectation value of important Pauli operators (having nonzero values) for the GHZ-cat can be directly expressed in terms of these coefficients:
\begin{eqnarray}
    &&\langle III\rangle=\langle ZZI\rangle=\langle ZIZ\rangle=\langle IZZ\rangle=\eta_1+\eta_4\nonumber \\
    &&\langle XXX\rangle=-\langle XYY\rangle=-\langle YXY\rangle=-\langle YYX\rangle=\eta_2+\eta_3.\nonumber \\
    \label{EQ_GHZdecayform_to_pauli}
\end{eqnarray}

For the W-cat, the state after decoherence has the form 
\begin{eqnarray}
    \rho &=& \eta_1|011\rangle \langle 011|+\eta_2|011\rangle \langle 101|+\eta_3|011\rangle \langle 110|\nonumber \\
    &&\eta_4|101\rangle \langle 011|+\eta_5|101\rangle \langle 101|+\eta_6|101\rangle \langle 110|\nonumber \\
    &&\eta_7|110\rangle \langle 011|+\eta_8|110\rangle \langle 101|+\eta_9|110\rangle \langle 110|.\label{EQ_Wstatedecayform}
\end{eqnarray}
The expectation value of important Pauli operators for the W-cat are expressed in terms of $\eta$s as 
\begin{eqnarray}
    &&\langle III\rangle=\langle ZZZ\rangle=\eta_1+\eta_5+\eta_9\nonumber \\
    &&\langle ZII\rangle=\langle IZZ\rangle=\eta_1-\eta_5-\eta_9\nonumber \\
    &&\langle IZI\rangle=\langle ZIZ\rangle=-\eta_1+\eta_5-\eta_9\nonumber \\
    &&\langle IIZ\rangle=\langle ZZI\rangle=-\eta_1-\eta_5+\eta_9\nonumber \\
    &&\langle XXI\rangle=\langle YYI\rangle=-\langle XXZ\rangle=-\langle YYZ\rangle=\eta_2+\eta_4\nonumber\\
    &&\langle XIX\rangle=\langle YIY\rangle=-\langle XZX\rangle=-\langle YZY\rangle=\eta_3+\eta_7\nonumber\\
    &&\langle IXX\rangle=\langle IYY\rangle=-\langle ZXX\rangle=-\langle ZYY\rangle=\eta_6+\eta_8.\nonumber \\
    \label{EQ_Wstatedecayform_to_pauli}
\end{eqnarray}

\subsection{Errors during state preparation}\label{App_EA_stateprep}

Here we will analyze the impact of each error channel independently on the expectation value of important Pauli operators for both the GHZ-cat and W-cat cases during state preparation.
The derivation procedure is as follows: apply the transformation of the corresponding error channel during the state preparation sequence and translate the impact from the oscillators' state to the expectation value of Pauli operators as described in Eqs.~(\ref{EQ_GHZdecayform_to_pauli}) and (\ref{EQ_Wstatedecayform_to_pauli}).
We will present the impact of decoherence as detrimental factors ($\le1$), multiplying the ideal expectation value of the Pauli operators.

\subsubsection{GHZ-cat}
The state preparation sequence for the GHZ-cat is illustrated in Fig.~2(b) in the main text. 
It includes a transmon rotation, an ECD, another transmon rotation, and a post selection.
The durations of the gates are $24$~ns, $1800$~ns, $24$~ns, and $1276$~ns, respectively.
The total duration of the ECD ($1800$~ns) comprises $1776$~ns of ECD main dynamics and $24$~ns of transmon correction pulse. 
As the duration for the transmon operations is much shorter than our system lifetimes, we can assume that these operations are perfect. 
Instead, we focus our attention on error contribution from the ECD main dynamics and post selection.

By considering only transmon energy decay, the detrimental factor for $\langle III\rangle$, $\langle ZZI\rangle$, $\langle ZIZ\rangle$, and $\langle IZZ\rangle$ is 
\begin{equation}
    1-\frac{t_1}{2T_1^{ge}},
\end{equation}
where $t_1=888$~ns is half the duration of the ECD main dynamics.
For $\langle XXX\rangle$, $\langle XYY\rangle$, $\langle YXY\rangle$, and $\langle YYX\rangle$, the detrimental factor is 
\begin{equation}
    1-\frac{t_1}{T_1^{ge}}-\frac{2t_2}{T_1^{ge}},
\end{equation} 
where $t_2=1276$~ns is the duration of the post selection pulse.
Note that, where applicable, we have assumed first order in $t_1/T_1^{ge}$ and $t_2/T_1^{ge}$ as these values are much smaller than $1$ in our case.

Next, by considering only transmon pure dephasing, the detrimental factor for $\langle III\rangle$, $\langle ZZI\rangle$, $\langle ZIZ\rangle$, and $\langle IZZ\rangle$ is $1$, whereas for $\langle XXX\rangle$, $\langle XYY\rangle$, $\langle YXY\rangle$, and $\langle YYX\rangle$ it is given by
\begin{equation}
    1-\frac{2t_1}{T_{\phi}^{ge}}.
\end{equation}

When only oscillators' photon loss is taken into account, the detrimental factor for $\langle III\rangle$, $\langle ZZI\rangle$, $\langle ZIZ\rangle$, and $\langle IZZ\rangle$ is $1$.  
For $\langle XXX\rangle$, $\langle XYY\rangle$, $\langle YXY\rangle$, and $\langle YYX\rangle$ it is given by
\begin{equation}
    \tilde f_1 \tilde f_2,\label{EQ_dfactorghzstateprepcavT1}
\end{equation}
where $\tilde f_1$ and $\tilde f_2$ denote the three mode decoherence factors (computed following Eq.~(\ref{EQ_3Mcoherence_decay_factor})) integrated during the ECD operation and post selection, respectively.

We present all the values of the detrimental factors computed with our experimental parameters in Table~\ref{tab:GHZ-cat_error_stateprep}. The Pauli operators $III$, $ZZI$, $ZIZ$ and $IZZ$, all of which are related to the diagonal elements in the computational space (or the population of the cat), experience negligible detrimental factors. On the other hand, the Pauli operators $XXX$, $XYY$, $YXY$ and $YYX$ experience more severe factors, mostly due to the oscillators' photon loss. This is because both $\tilde f_1$ and $\tilde f_2$ decay exponentially with the size of the coherent separation during the ECD and post selection, respectively. 

Note that the detrimental factors $\eta_1$ and $\eta_4$ translate to the population that remains within the computational space $p_S$ (after the considered decoherence): $\eta_1+\eta_4=p_{S}$.
In our case, with our experimental parameters, $p_S\approx 0.99$.
This means that the probability of leakage out of the computational space $\approx 1$\%, which supports the use of the subspace tomography in our work.

\begin{table}[h]
\centering
\begin{tabular}{c|c|c|c|c|c|c|c|c|}
 & III & ZZI & ZIZ & IZZ & XXX & XYY & YXY & YYX\\ \hline
q $T_1$ & $0.99$ & $0.99$ & $0.99$ & $0.99$ & $0.93$ & $0.93$ & $0.93$ & $0.93$ \\ \hline
q $T_{\phi}$ & $1$ & $1$ & $1$ & $1$ & $0.96$ & $0.96$ & $0.96$ & $0.96$ \\ \hline
osc $T_1$ & $1$ & $1$ & $1$ & $1$ & $0.72$ & $0.72$ & $0.72$ & $0.72$ \\ \hline
\end{tabular}
\caption{\textbf{Summary of detrimental factors of the GHZ-cat during state preparation}. The expectation value of Pauli operators are listed in the column. The rows q $T_1$, q $T_{\phi}$, and osc $T_1$ denote the error channel for transmon energy decay, transmon pure dephasing, and oscillators' photon loss, respectively.}
\label{tab:GHZ-cat_error_stateprep}
\end{table}

\subsubsection{W-cat}

The W-cat state preparation sequence is presented in Fig.~3(c) in the main text. 
It features an ECD, a UECD, a post selection, transmon rotations and oscillator displacements.
Similar to the treatment of the GHZ-cat above, we will ignore decoherence effects for fast operations such as transmon rotations and oscillator displacements.
As such, we will focus our attention on error contribution from the ECD, UECD, and post selection.
The durations for the main dynamics of ECD and UECD are $1776$~ns and $1800$~ns, respectively, while the post selection pulse is $1012$~ns long. 

By considering only transmon energy decay, the detrimental factor for $\langle III \rangle$ and $\langle ZZZ \rangle$ is 
\begin{equation}
    1-\frac{t_1}{T_{1}^{ge}}-\frac{t_1^{\prime}}{3T_{1}^{ge}}-\frac{t_1^{\prime}}{3T_{1}^{ef}},
\end{equation}
where $t_1=888$~ns is half of the duration of the ECD main dynamics and $t_1^{\prime}=900$~ns is half that of the UECD main dynamics.
The factor for $\langle ZII \rangle$ and $\langle IZZ \rangle$ is 
\begin{equation}
    1-\frac{t_1}{T_{1}^{ge}}-\frac{t_1^{\prime}}{T_{1}^{ge}}-\frac{t_1^{\prime}}{T_{1}^{ef}}+\frac{t_2^{\prime}} {T_{1}^{ge}},
\end{equation}
where $t_2^{\prime}=1012$~ns is the duration of the post selection pulse.
The factor for $\langle IZI \rangle$ and $\langle ZIZ \rangle$ is
\begin{equation}
    1-\frac{t_1}{T_{1}^{ge}}-\frac{t_1^{\prime}}{T_{1}^{ge}}+\frac{t_1^{\prime}}{T_{1}^{ef}}-\frac{2t_2^{\prime}} {T_{1}^{ge}};
\end{equation}
for $\langle IIZ \rangle$ and $\langle ZZI \rangle$ is
\begin{equation}
    1-\frac{t_1}{T_{1}^{ge}}+\frac{t_1^{\prime}}{T_{1}^{ge}}-\frac{t_1^{\prime}}{T_{1}^{ef}}+\frac{t_2^{\prime}} {T_{1}^{ge}};
\end{equation}
for $\langle XXI \rangle$, $\langle YYI \rangle$, $\langle XXZ \rangle$ and $\langle YYZ \rangle$ is
\begin{equation}
    1-\frac{t_1}{T_{1}^{ge}}-\frac{t_1^{\prime}}{T_{1}^{ef}}-\frac{2t_2^{\prime}} {T_{1}^{ge}};
\end{equation}
for $\langle XIX \rangle$, $\langle YIY \rangle$, $\langle XZX \rangle$ and $\langle YZY \rangle$ is
\begin{equation}
    1-\frac{t_1}{T_{1}^{ge}}-\frac{t_1^{\prime}}{T_{1}^{ge}}-\frac{t_1^{\prime}}{2T_{1}^{ef}}-\frac{t_2^{\prime}} {2T_{1}^{ge}};
\end{equation}
and lastly for $\langle IXX \rangle$, $\langle IYY \rangle$, $\langle ZXX \rangle$ and $\langle ZYY \rangle$ is
\begin{equation}
    1-\frac{t_1}{T_{1}^{ge}}-\frac{t_1^{\prime}}{T_{1}^{ge}}-\frac{t_1^{\prime}}{2T_{1}^{ef}}-\frac{2t_2^{\prime}} {T_{1}^{ge}}.
\end{equation}
Note that we have assumed first order in $t_1/T_1^{ge}$, $t_1^{\prime}/T_1^{ge}$, $t_1^{\prime}/T_1^{ef}$, and $t_2^{\prime}/T_1^{ge}$.

Next, when only transmon dephasing is taken into account, the factor for $\langle III \rangle$, $\langle ZII \rangle$, $\langle IZI \rangle$, $\langle IIZ \rangle$, $\langle ZZI \rangle$, $\langle ZIZ \rangle$, $\langle IZZ \rangle$, and $\langle ZZZ \rangle$ is 1.
The factor for $\langle XXI \rangle$, $\langle YYI \rangle$, $\langle XXZ \rangle$, and $\langle YYZ \rangle$ is given by
\begin{equation}
    e^{-\frac{2t_1}{T_{\phi}^{ge}}}e^{-\frac{2t_1^{\prime}}{T_{\phi}^{gf}}},
\end{equation}
where $T_{\phi}^{gf}\equiv(1/T_{\phi}^{ef}-1/T_{\phi}^{ge})^{-1}$ is the induced dephasing lifetime in the g-f space.
The factor for $\langle XIX \rangle$, $\langle YIY \rangle$, $\langle XZX \rangle$, and $\langle YZY \rangle$ is
\begin{equation}
    e^{-\frac{2t_1}{T_{\phi}^{ge}}}e^{-\frac{t_1^{\prime}}{T_{\phi}^{ge}}}e^{-\frac{t_1^{\prime}}{T_{\phi}^{ef}}},
\end{equation}
and for $\langle IXX \rangle$, $\langle IYY \rangle$, $\langle ZXX \rangle$, and $\langle ZYY \rangle$ it is given by
\begin{equation}
    e^{-\frac{t_1^{\prime}}{T_{\phi}^{ge}}}e^{-\frac{t_1^{\prime}}{T_{\phi}^{ef}}}.
\end{equation}

Finally, considering only oscillators' photon loss, the detrimental factors eventually transfer to the oscillators' computational space as expressed in Eq.~(\ref{EQ_Wstatedecayform}) where $\eta_1=\eta_5=\eta_9=1/3$. Other detrimental factors are computed following Eq.~(\ref{EQ_3Mcoherence_decay_factor}) integrated during the ECD, UCED, and post selection pulse.

With our experimental parameters, all the detrimental parameters are computed and summarized in Table~\ref{tab:W-cat_error_stateprep}. Similar to the case of the GHZ-cat, the Pauli operators related to the diagonal elements (in the computational space): $III$, $ZII$, $IZI$, $IIZ$, $ZZI$, $ZIZ$, $IZZ$, and $ZZZ$ have detrimental factors close to unity. On the other hand, the two and three mode correlations experience more impact.
The major contributing factor being the oscillators' photon loss on the Pauli operators: $XXI$, $YYI$, $XXZ$, $YYZ$, $XIX$, $YIY$, $XZX$, and $YZY$. 
Note that the relatively more moderate factors on Bob-Charlie correlations $IXX$, $IYY$, $ZXX$, and $ZYY$ cause the system to have largest residual bipartite entanglement between the two parties, higher than Alice-Bob and Alice-Charlie bipartitions. 

Similar to the case of GHZ-cat, the detrimental factors $\eta_1$, $\eta_5$ and $\eta_9$ translate to the population that remains within the computational space $p_S$ (after the considered decoherence): $\eta_1+\eta_5+\eta_9=p_{S}$.
In our case, with our experimental parameters, $p_S\approx 0.96$.
This means that the probability of leakage out of the computational space $\approx 4$\%.

\begin{table*}[t]
\centering
\begin{tabular}{c|c|c|c|c|c|c|c|c|c|c|c|c|c|c|c|c|c|c|c|c|}
 & III & ZII & IZI & IIZ & ZZI & ZIZ & IZZ & ZZZ
 & XXI & YYI & XXZ & YYZ
 & XIX & YIY & XZX & YZY
 & IXX & IYY & ZXX & ZYY\\ \hline
q $T_1$ & $0.96$ & $0.94$ & $0.97$ & $0.98$ & $0.98$ & $0.97$ & $0.94$ & $0.96$ &
$0.90$ & $0.90$ & $0.90$ & $0.90$ &
$0.93$ & $0.93$ & $0.93$ & $0.93$ &
$0.90$ & $0.90$ & $0.90$ & $0.90$ \\ \hline
q $T_{\phi}$ & $1$ & $1$ & $1$ & $1$ & $1$ & $1$ & $1$ & $1$ &
$0.96$ & $0.96$ & $0.96$ & $0.96$ &
$0.92$ & $0.92$ & $0.92$ & $0.92$ &
$0.96$ & $0.96$ & $0.96$ & $0.96$ \\ \hline
osc $T_{1}$ & $1$ & $1$ & $1$ & $1$ & $1$ & $1$ & $1$ & $1$ &
$0.56$ & $0.56$ & $0.56$ & $0.56$ &
$0.57$ & $0.57$ & $0.57$ & $0.57$ &
$0.82$ & $0.82$ & $0.82$ & $0.82$ \\ \hline
\end{tabular}
\caption{\textbf{Summary of detrimental factors of the W-cat during state preparation}. The expectation value of Pauli operators are listed in the column. The rows q $T_1$, q $T_{\phi}$, and osc $T_1$ denote the error channel for transmon energy decay, transmon pure dephasing, and oscillators' photon loss, respectively.}
\label{tab:W-cat_error_stateprep}
\end{table*}

\subsection{Errors during state characterization}\label{App_EA_statechar}

Here we present errors during state characterization via CF measurements, the sequence of which is presented in Fig.~2(b) in the main text.

Let us first assume arbitrary three-mode state $\rho$ and the transmon in the ground state $|g\rangle$ before the CF sequence. In the ideal case, following the sequence, one gets the real and imaginary parts of the CF from the last transmon measurement as follows:
\begin{equation}
    1-2p_g=\text{Re}(\langle D(2\alpha)\rangle),
\end{equation}
where $p_g$ is the probability of the transmon being in the ground state, if the last transmon rotation is performed along the y-axis and 
\begin{equation}
    2p_g-1=\text{Im}(\langle D(2\alpha)\rangle)
\end{equation}
if the rotation is performed along the x-axis.
In the presence of an error channel, as discussed previously, we will only apply the error channel during the ECD as the durations for the transmon rotations are much shorter.

In the presence of transmon energy decay alone, the corresponding expressions for the CF are modified as
\begin{eqnarray}
    1-2p_g&=&\text{Re}(\langle D(2\alpha)\rangle)e^{-\frac{t_1}{T_1^{ge}}}\nonumber \\
    2p_g-1&=&\text{Im}(\langle D(2\alpha)\rangle)e^{-\frac{t_1}{T_1^{ge}}},\label{EQ_CFtransmondecay}
\end{eqnarray}
where $t_1=888$~ns is half the duration of the ECD main dynamics. Both the real and imaginary parts of the CF get a detrimental factor of $\exp{(-t_1/T_1^{ge})}$, which in our case is $\approx0.98$. 
Note that this detrimental factor is independent of the oscillators' state and CF displacement point $\alpha$.

Similarly, considering only transmon pure dephasing, the expressions read
\begin{eqnarray}
    1-2p_g&=&\text{Re}(\langle D(2\alpha)\rangle)e^{-\frac{2t_1}{T_{\phi}^{ge}}}\nonumber \\
    2p_g-1&=&\text{Im}(\langle D(2\alpha)\rangle)e^{-\frac{2t_1}{T_{\phi}^{ge}}},\label{EQ_CFtransmondephasing}
\end{eqnarray}
where now the universal detrimental factor is $\exp{(-2t_1/T_{\phi}^{ge})}$, which is $\approx 0.96$ in our case.

For the case of oscillators' photon loss, the detrimental factors are more involved. They depend on the CF displacement point and oscillators' state being characterized. Before analyzing specific CF points and oscillators' states, let us first assume a general oscillators' state $\rho$ and derive the CF expressions.

Consider the measurement of $\text{Re}(\langle D(\vec \beta\rangle)$, where the general joint displacement operator is defined as $D(\vec \beta) \equiv D_A(\beta_A)\otimes D_B(\beta_B)\otimes D_C(\beta_C)$.
The sequence starts with transmon $R_y(\pi/2)$ rotation in the g-e subspace, after which the transmon-oscillators state is expressed as $0.5(|g\rangle\langle g|+|g\rangle\langle e|+|e\rangle\langle g|+|e\rangle\langle e|)\rho$.
Next, we apply $ECD(\vec \beta)=|e\rangle \langle g| D(\vec \beta/2)+|g\rangle \langle e| D(\text{-}\vec \beta/2)$.
The ECD main dynamics can be divided into 3 parts. 
Part 1 effectively applies an operation $|g\rangle \langle g|\tilde D(\vec \beta/4)+|e\rangle \langle e|\tilde D(\text{-}\vec \beta/4)$. 
The tilde, for example in $\tilde D(\vec \beta/4)$, denotes the fact that the actual operation is not a displacement $D(\vec \beta/4)$. 
It consists of a displacement in the perpendicular direction, a phase space rotation (wait), and a displacement back such that the trajectory ends up at $\vec \beta/4$. 
For exemplary illustration, we refer to Fig.~2(a) in the main text, where this operation for $\vec \beta=(3,3,3)$ is represented by the solid trajectories in the bottom right corner. 
After the operation, we can write the state as 
\begin{eqnarray}
    \rho_1 &=&\frac{1}{2}(|g\rangle \langle g|\tilde D(\vec \beta/4)\rho \tilde D(\text{-}\vec \beta/4)+|g\rangle \langle e|\tilde D(\vec \beta/4)\rho \tilde D(\vec \beta/4)\nonumber \\
    &&+|e\rangle \langle g|\tilde D(\text{-}\vec \beta/4)\rho \tilde D(\text{-}\vec \beta/4) +|e\rangle \langle e|\tilde D(\text{-}\vec \beta/4)\rho \tilde D(\vec \beta/4)).\nonumber \\
\end{eqnarray}
Part 2 is a fast transmon flip $|g\rangle \leftrightarrow |e\rangle$, where effects of decoherence are ignored. Now the state simply reads
\begin{eqnarray}
    \rho_2 &=&\frac{1}{2}(|e\rangle \langle e|\tilde D(\vec \beta/4)\rho \tilde D(\text{-}\vec \beta/4)+|e\rangle \langle g|\tilde D(\vec \beta/4)\rho \tilde D(\vec \beta/4)\nonumber \\
    &&+|g\rangle \langle e|\tilde D(\text{-}\vec \beta/4)\rho \tilde D(\text{-}\vec \beta/4) +|g\rangle \langle g|\tilde D(\text{-}\vec \beta/4)\rho \tilde D(\vec \beta/4)).\nonumber \\
\end{eqnarray}
Finally, part 3 effectively applies an operation $|g\rangle \langle g|\tilde {\tilde D}(\text{-}\vec \beta/4)+|e\rangle \langle e|\tilde {\tilde D}(\vec \beta/4)$.
Similar to part 1, this operation for $\vec \beta=(3,3,3)$ is represented by solid trajectories in the top left corner of Fig.~2(a) in the main text and the double tilde notation is to indicate that it is a separate trajectory compared to the ones in part 1.
After the operation, we write the state as
\begin{equation}
    \rho_3 = \frac{1}{2}(|e\rangle \langle e|\fbox{1}+|e\rangle \langle g|\fbox{3}+|g\rangle \langle e|\fbox{4}+|g\rangle \langle g|\fbox{2}),
\end{equation}
where the numbered boxes are defined as follows
\begin{eqnarray}
    \fbox{1} &=&\tilde {\tilde D}(\vec \beta/4) \tilde D(\vec \beta/4)\rho \tilde D(\text{-}\vec \beta/4) \tilde {\tilde D}(\text{-}\vec \beta/4)\nonumber \\
    \fbox{3}&=&\tilde {\tilde D}(\vec \beta/4)\tilde D(\vec \beta/4)\rho \tilde D(\vec \beta/4)\tilde {\tilde D}(\vec \beta/4)\nonumber \\
    \fbox{4}&=&\tilde {\tilde D}(\text{-}\vec \beta/4) \tilde D(\text{-}\vec \beta/4)\rho \tilde D(\text{-}\vec \beta/4) \tilde {\tilde D}(\text{-}\vec \beta/4)\nonumber \\ \fbox{2}&=&\tilde {\tilde D}(\text{-}\vec \beta/4) \tilde D(\text{-}\vec \beta/4)\rho \tilde D(\vec \beta/4)\tilde {\tilde D}(\vec \beta/4)). \label{EQ_box1234}
\end{eqnarray}
After the last transmon rotation $R_y(\pi/2)$ in the g-e subspace, the transmon's probability in the ground state reads
\begin{equation}
    p_g=\frac{1}{4}(\text{tr}(\fbox{1})+\text{tr}(\fbox{2})-\text{tr}(\fbox{3})-\text{tr}(\fbox{4})),\label{EQ_pgrealCF}
\end{equation}
from which the real part of the CF, $1-2p_g$, is computed.

On the other hand, the measurement of $\text{Im}(\langle D(\vec \beta\rangle)$ uses $R_x(\pi/2)$ as the last transmon rotation, after which the probability of the transmon in the ground state reads 
\begin{equation}
    p_g=\frac{1}{4}(\text{tr}(\fbox{1})+\text{tr}(\fbox{2})-i\text{tr}(\fbox{3})+i\text{tr}(\fbox{4})),\label{EQ_pgimagCF}
\end{equation}
which gives us the imaginary part of the CF as $2p_g-1$.

Now we analyze the expectation value of Pauli operators for the tripartite states. 
In this section, we will assume perfect GHZ-cat and W-cat states before the CF sequence for simplicity.
The results can be applied to states that have been affected by decoherence in the state preparation stage, i.e., the GHZ-cat and W-cat states given in Eq.~(\ref{EQ_GHZdecayform}) and Eq.~(\ref{EQ_Wstatedecayform}), respectively.

For the GHZ-cat, $\langle III\rangle=\text{Re}(\langle D(0,0,0)\rangle)$. In this case, it is clear from Eq.~(\ref{EQ_box1234}) that $\fbox{1}=\fbox{2}=\fbox{3}=\fbox{4}$ and the GHZ-state $\rho$ would simply decay in amplitude (negligible) and off-diagonal elements (which can be computed via Eq.~(\ref{EQ_3Mcoherence_decay_factor})) during the ECD. However, only the trace would contribute to the CF in the end. In this case, the trace of all the terms equal 1. Therefore, $p_g=0$ in Eq.~(\ref{EQ_pgrealCF}), and hence, $\text{Re}(\langle D(0,0,0)\rangle)\equiv 1-2p_g=1$. In other words, the detrimental factor is 1.

A different case is given by $\langle XXX \rangle=2(\text{Re}(\langle D(2\alpha,2\alpha,2\alpha)\rangle)+\text{Re}(\langle D(2\alpha,2\alpha,\text{-}2\alpha)\rangle)+\text{Re}(\langle D(2\alpha,\text{-}2\alpha,2\alpha)\rangle)+\text{Re}(\langle D(\text{-}2\alpha,2\alpha,2\alpha)\rangle))$. We will take a closer look at $\text{Re}(\langle D(2\alpha,2\alpha,2\alpha)\rangle)$ as it is the only nonzero term for the GHZ-cat state.
This requires performing the ECD operation $ECD(2\alpha,2\alpha,2\alpha)$, in which one can see that $\fbox{1}$ and $\fbox{2}$ represent a GHZ-cat being displaced by $\vec \alpha$ and $\text{-}\vec \alpha$, respectively. While the states decay during the process, their traces remain 1.
On the other hand, a nonzero contributing term from $\fbox{3}$ comes from the off-diagonal element of the GHZ-cat state: $\tilde {\tilde D}(\vec \alpha/2)\tilde D(\vec \alpha/2)(1/2)|\text{-}\alpha,\text{-}\alpha,\text{-}\alpha\rangle \langle \alpha,\alpha,\alpha | \tilde D(\vec \alpha/2)\tilde {\tilde D}(\vec \alpha/2)\rightarrow(1/2)|0,0,0\rangle\langle 0,0,0|\tilde f_1$, where $\tilde f_1$ is a coherence decay factor computed following Eq.~(\ref{EQ_3Mcoherence_decay_factor}), integrated during the ECD trajectory from $|\text{-}\alpha,\text{-}\alpha,\text{-}\alpha\rangle \langle \alpha,\alpha,\alpha |$ to $|0,0,0\rangle \langle0,0,0|$. This detrimental factor $\tilde f_1$ is the same as that described in Eq.~(\ref{EQ_dfactorghzstateprepcavT1}), which was computed for the opposite trajectory from $|0,0,0\rangle \langle0,0,0|$ to $|\text{-}\alpha,\text{-}\alpha,\text{-}\alpha\rangle \langle \alpha,\alpha,\alpha |$.
Therefore, we have $\text{tr}(\fbox{3})=\tilde f_1/2$.
By following the same method, we also have $\text{tr}(\fbox{4})=\tilde f_1/2$, which finally gives $\text{Re}(\langle D(2\alpha,2\alpha,2\alpha)\rangle)=\tilde f_1/2$, and hence $\langle XXX \rangle = \tilde f_1$. The factor $\tilde f_1$ is directly the detrimental factor as the ideal expectation value of this Pauli operator is 1.

We apply the same method to evaluate $\langle XYY\rangle$, $\langle YXY\rangle$, $\langle YYX\rangle$, $\langle ZZI\rangle$, $\langle ZIZ\rangle$, and $\langle IZZ\rangle$. The detrimental factors due to oscillators' photon loss are $\tilde f_1$, $\tilde f_1$, $\tilde f_1$, 1, 1, and 1, respectively. With our experimental parameters, we have $\tilde f_1\approx 0.90$.

Similarly, we performed the same analysis for the case of the W-cat state. We found that $\langle III \rangle$, $\langle ZII \rangle$, $\langle IZI \rangle$, $\langle IIZ \rangle$, $\langle ZZI \rangle$, $\langle ZIZ \rangle$, $\langle IZZ \rangle$, and $\langle ZZZ \rangle$ all have detrimental factors of 1. 
By contrast, $\langle XXI \rangle$, $\langle YYI \rangle$, $\langle XXZ \rangle$, and $\langle YYZ \rangle$ have detrimental factors of $\tilde f_{1(AB)}$, where it is computed similarly to $\tilde f_1$ for only mode $A$ and $B$ as no displacement is applied on mode $C$.
Next, $\langle XIX \rangle$, $\langle YIY \rangle$, $\langle XZX \rangle$, and $\langle YZY \rangle$ have detrimental factors of $\tilde f_{1(AC)}$ considering only the trajectory for mode $A$ and $C$, while $\langle IXX \rangle$, $\langle IYY \rangle$, $\langle ZXX \rangle$, and $\langle ZYY \rangle$ all have detrimental factors of $\tilde f_{1(BC)}$ considering only the trajectory for mode $B$ and $C$.
With our experimental parameters the values for $\tilde f_{1(AB)}$, $\tilde f_{1(AC)}$ and $\tilde f_{1(BC)}$ are respectively $\approx0.92$, $\approx0.95$, and $\approx0.92$.

\subsection{All errors vs experimental data}\label{App_EA_allerror}

So far, we have done the error analysis separately in the presence of transmon energy decay, transmon dephasing, and oscillators' photon loss during both the state preparation and characterization. 
The transmon decoherence results in detrimental factors that are relatively minor.
The oscillators' photon loss results in decay of the coherent state amplitudes, which we ignore since the effect is negligible, and coherence decay of the ``off-diagonal" elements when the state is in a superposition of coherent states, whose contribution is major in our work.

\begin{figure*}
    \centering
    \includegraphics[]{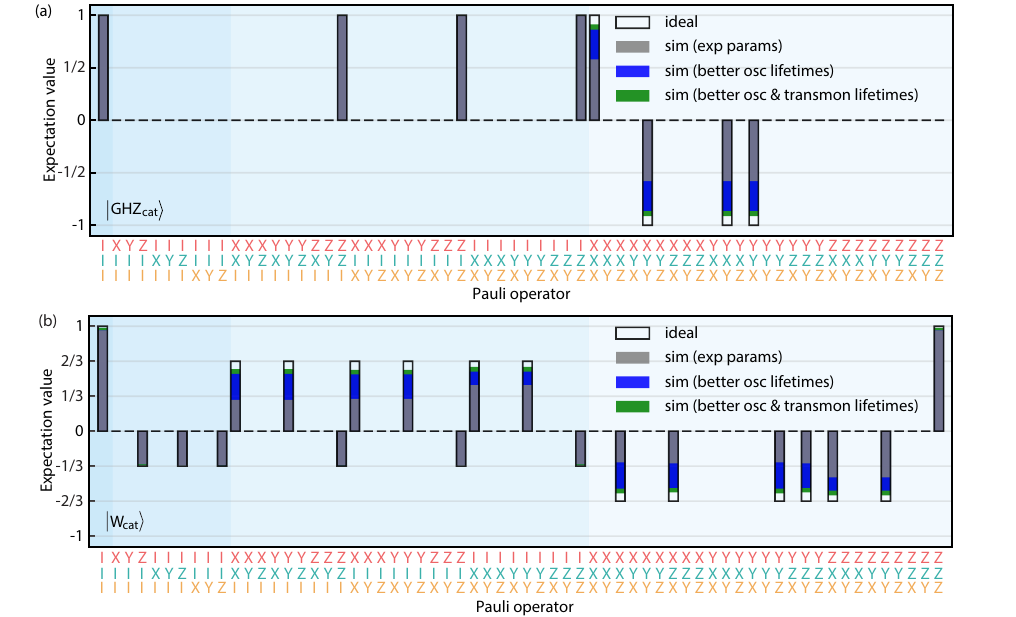}
    \caption{\textbf{Expectation values of Pauli operators for (a) GHZ-cat and (b) W-cat states}. In both panels white bars represent ideal case, gray bars the simulation with our experimental parameters, blue bars when the oscillators' lifetimes are increase to $1$~ms, and green bars when the oscillators' lifetimes are $1$~ms and the transmon lifetimes are $\sim 100\:\mu$s.}
    \label{fig:spaulis}
\end{figure*}

In the attempt to combine all errors to explain experimental data, we note the following observations.
The contributions from transmon energy decay and dephasing can be approximated to first order, in which the detrimental factors multiply both during the state preparation and characterization.
We can assume oscillators' photon loss gives the main detrimental factors to the density matrix, which affect the ``off-diagonal" terms during the protocol and the detrimental factors would multiply with first-order detrimental factors from both transmon energy decay and dephasing (both during the state preparation and characterization).
As we noted previously, the GHZ-cat and W-cat states after the state preparation with decoherence would have the forms given by Eq.~(\ref{EQ_GHZdecayform}) and Eq.~(\ref{EQ_Wstatedecayform}), respectively. 
We only note the elements in this subspace as the ones outside the subspace would not contribute to the expectation value of the Pauli operators.
As such, the analysis that we performed for errors during the state characterization can be easily extended to the case starting with these imperfect forms for the GHZ-cat and W-cat states.
As a result, we can multiply the detrimental factors from the state preparation and characterization for each Pauli operator.
In the end, we would have a total detrimental factor for each expectation value of Pauli operator for both the GHZ-cat and W-cat states.

For the experimental data, labeled ``exp" in Fig.~2 and Fig.~4 in the main text, the data is calibrated with vacuum correction, which is a standard practice in the community. 
This calibration involves starting the sequence with vacuum for all oscillators and directly performing the CF measurements. 
In this case, the error channels affecting the measurement would be the transmon energy decay and dephasing with a universal factor $e^{-\frac{t_1}{T_1^{ge}}}e^{-\frac{2t_1}{T_{\phi}^{ge}}}$ (see Eqs.~(\ref{EQ_CFtransmondecay}) and (\ref{EQ_CFtransmondephasing})). 
To be consistent with the calibrated experimental data, we will ignore this factor in our calculation. 
After accumulating the remaining detrimental factors and multiplying them with the ideal expectation values of the corresponding Pauli operators, the results are plotted as gray bars in Fig.~2 in the main text for the GHZ-cat state and gray bars in Fig.~4 in the main text for the W-cat state.
In both cases, we see that the values are in good agreement with the experimental data.
This agreement justifies the error analysis that we derived, which can be further used for example to demonstrate the capability of the protocols should the systems' lifetimes are improved, which we discuss below, and to devise a control experiment that can further calibrates the errors during the state characterization, which we discuss in the next section.

For the case where the oscillators' lifetimes are improved to $T_{A1}=T_{B1}=T_{C1}=1$~ms, we present the expectation values of the Pauli operators of the GHZ-cat in Fig.~\ref{fig:spaulis}(a) (blue bars). If one further improves the transmon lifetimes to $T_1^{ge}=100\:\:\mu$s, $T_{\phi}^{ge}=80\:\:\mu$s, $T_1^{ef}=90\:\:\mu$s, and $T_{\phi}^{ef}=70\:\:\mu$s, the results are shown as green bars.
Similarly, the case of the W-cat state is presented in Fig.~\ref{fig:spaulis}(b).
Post processing the expectation values of the Pauli operators from our error models results in GHZ-cat fidelities: $0.78\pm0.02$ (our experimental parameters); $0.93\pm 0.03$ (improved oscillators' lifetime to $1$~ms); and $0.95\pm 0.03$ (further improvement on transmon lifetimes to $\sim100\:\mu$s).
For the W-cat state, we obtain fidelities: $0.67\pm0.02$ (our experimental parameters); $0.88\pm 0.03$ (improved oscillators' lifetime to $1$~ms); and $0.93\pm 0.03$ (further improvement on transmon lifetimes to $\sim100\:\mu$s).
We note that the fidelities obtained from the error models using our experimental parameters $0.78\pm 0.02$ (GHZ-cat) and $0.67\pm 0.02$ (W-cat) are in good agreement with the ones obtained by post-processing the measured Pauli operators, labeled ``exp" in Fig.~2 and 4 in the main text, which are $0.79\pm 0.02$ (GHZ-cat) and $0.64\pm 0.02$ (W-cat), respectively.

\subsection{Control experiments}\label{App_EA_controlexp}

Our derivations for the errors in an \textit{independent} way, both among the error channels and between the state preparation and characterization, makes it easy to see the contribution (in a form of detrimental factor) from each error channel to each process. 
Additionally, this motivates us to devise a control experiment to calibrate the errors during the state characterization, such that we can assess the quality of the prepared states with errors solely coming from the state preparation protocol. 

For the state characterization, the detrimental factors from transmon energy decay and dephasing have been calibrated out via the vacuum calibration experiment described in the previous section.
Here, we will describe experiments to calibrate the remaining factors $\tilde f_1$, $\tilde f_{1(AB)}$, $\tilde f_{1(AC)}$, and $\tilde f_{1(BC)}$ due to oscillators' photon loss.

The control experiment is essentially given by the cat-and-back experiment, the sequence of which is given in Fig.~\ref{fig:s3}(a).
Let us consider the case where $\beta=2\alpha$.
In the ideal case, the transmon rotation $R_x(\pi/2)$ and ECD operation $ECD(2\alpha,2\alpha,2\alpha)$ prepares a superposition state $(|e\rangle |\alpha,\alpha,\alpha\rangle-i|g\rangle|\text{-}\alpha,\text{-}\alpha,\text{-}\alpha\rangle )/\sqrt{2}$, where we have ignored the geometric phase.
Then the rest of the operations ($R_x(\pi)$, $ECD(\text{-}2\alpha,\text{-}2\alpha,\text{-}2\alpha)$, and $R_x(\pi/2)$) bring the state to
$|g\rangle |0,0,0\rangle$.
In the presence of decoherence, following our error models, we found that the transmon measurement yields 
\begin{equation}
    \langle \sigma_z\rangle=2p_g-1=(\tilde f_1e^{-\frac{2t_1}{T_{\phi}^{ge}}}e^{-\frac{t_1}{T_1^{ge}}})^2,\label{EQ_controlexpalpha}
\end{equation}
which encodes double the detrimental factor of a single ECD operation.
If we now consider the case where $\beta=0$, the transmon measurement would result in 
\begin{equation}
    \langle \sigma_z \rangle=2p_g-1=(e^{-\frac{2t_1}{T_{\phi}^{ge}}}e^{-\frac{t_1}{T_1^{ge}}})^2.\label{EQ_controlexp0} 
\end{equation}
This, together with Eq.~(\ref{EQ_controlexpalpha}), would allow us to compute $\tilde f_1$. 
For variable $\beta$, the three-mode cat-and-back experiment results in oscillation in $\langle \sigma_z\rangle$, shown as brown circles in Fig.~\ref{fig:s3}(b) (right panel), which is due to the geometric phase. The purely decaying part of the amplitude that is associated Eq.~(\ref{EQ_controlexpalpha}) and Eq.~(\ref{EQ_controlexp0}) at $\beta=2\alpha$ and $\beta=0$, respectively, is obtained by fitting the envelope of the oscillating experimental data, the result of which is plotted as dashed curve.
The analysis using the three-mode cat-and-back experiment reveals $\tilde f_1\approx 0.91$, which we use to calibrate the affected expectation value of the Pauli operators for the GHZ-cat state in Fig.~2 in the main text.
We note that the obtained factor is very close to $0.90$, computed independently using our error model. 

Alternatively, one can perform the cat-and-back experiment with a single mode ECD. For example, one simply replaces the two ECD operations in Fig.~\ref{fig:s3}(a) with $ECD(\beta,0,0)$ and $ECD(\text{-}\beta,0,0)$, respectively. An exemplary experimental data for this is presented in Fig.~\ref{fig:s3}(b) (left panel).
Similar to the above analysis, this can be used to retrieve the detrimental factor $\tilde f_{1(A)}$, resulting from oscillator's $A$ photon loss affecting its state during the ECDs.
The same analysis can be done for mode $B$, from which we get $\tilde f_{1(B)}$. 
It is clear, following the discussion of Eq.~(\ref{EQ_3Mcoherence_decay_factor}), that the detrimental factor $\tilde f_{1(AB)}=\tilde f_{1(A)}\tilde f_{1(B)}$. 
We use this technique to independently retrieve $\tilde f_{1(AB)}$, $\tilde f_{1(AC)}$, and $\tilde f_{1(BC)}$ required to calibrate the affected expectation value of Pauli operators for the W-cat state in Fig.~4 in the main text.


\section{Data and codes availability}

The data and codes that support our findings are openly available on GitHub: \url{https://github.com/tkrisnanda/GHZ_W_cat}.

\bibliography{refs} 

\end{document}